\documentclass{aa}  
\usepackage{graphicx}
\usepackage{cancel}
\usepackage[colorlinks,allcolors=blue]{hyperref}
\bibpunct{(}{)}{;}{a}{}{,}
\usepackage{natbib}
\usepackage[varg]{txfonts}
\usepackage{tabularx} 
\usepackage{graphicx}
\usepackage{txfonts}
\def\apjl{\rm ApJL}
\def\apjs{\rm ApJS}

\def\aap{\rm A\&A}

\begin{document}

   \title{Instabilities at recollimation shocks in MHD jets}

 \author{S. Boula
          \inst{1}\orcid{0000-0001-7905-6928}, F. Tavecchio\inst{1}\orcid{0000-0003-0256-0995}, G. Bodo \inst{2}\orcid{0000-0002-9265-4081}, N. Vlahakis\inst{3}\orcid{0000-0002-8913-5176}, P. Coppi\inst{4}\orcid{0000-0001-9604-2325}, A. Costa\inst{2}\orcid{0009-0006-4561-4446}, 
          \and
         A.Sciaccaluga \inst{1}\orcid{0000-0001-6181-839X}
        }

   \institute{INAF – Osservatorio Astronomico di Brera, Via E. Bianchi 46, I-23807 Merate, Italy\\
              \email{styliani.boula@inaf.it}
         \and
            INAF, Osservatorio Astrofisico di Torino, Strada Osservatorio 20, I-10025 Pino Torinese, Italy
\and 
Department of Physics, National and Kapodistrian University of Athens, University Campus, Zografos, GR-157 84 Athens, Greece
\and 
Department of Astronomy, Yale University, PO Box 208101, New Haven, CT 06520-8101, USA
         }


 
  \abstract
{The internal structure and stability of relativistic jets from active galactic nuclei (AGN) still presents open questions relevant to high-energy astrophysics, with recollimation shocks often invoked to explain the jet morphology, particle acceleration, and variability. Yet, the role of instabilities triggered downstream of these shocks is not fully understood, particularly in magnetized jets.}
 {We aim to investigate how jet magnetization and other physical parameters influence the development of instabilities beyond the first recollimation shock. In particular, we focus on identifying the conditions under which the centrifugal instability (CFI) is effective, and how it affects the jet propagation and internal dynamics.}
 {We perform high-resolution 2D and 3D simulations using the relativistic magnetohydrodynamics code \texttt{PLUTO}. The jets are initialized with a conical geometry and propagate into an ambient medium, and we follow by axisymmetric simulations how they evolve towards a steady-state. In two-dimensions we explore a range of magnetizations (from $0$ to $1$), pressure contrasts, and inertia ratios to characterize the formation and evolution of recollimation shocks. The results are further evaluated using linear stability analysis to assess the growth and suppression of CFI. Finally, we perform 3D simulations of unstable and stable jets.}
 {We discuss how the different parameters of the axisymmetric steady solutions influence the location and strength of recollimation. We find that, even in moderately magnetized jets, $\sigma=0.1$, the CFI can still develop under suitable local conditions and disrupt the jet structure. This instability is governed by the jet radius, curvature, Lorentz factor, and magnetization, and is not always predictable from injection conditions. While magnetization can delay or locally suppress instability growth, it does not guarantee long-term jet stability. Our 3D results highlight the limitations of 2D models in capturing non-axisymmetric and nonlinear effects, and underline the complex interplay between magnetic confinement and destabilizing mechanisms. These findings have implications for interpreting variability, knot formation, and polarization structure in AGN jets.}
 {}
   \keywords{Magnetohydrodynamics (MHD)-- Instabilities --
               Shock waves --
                turbulence -- Galaxies: active
               }
 \titlerunning{Recollimation shocks and MHD jets stabilization}
 
   \authorrunning{Boula et al.}
   \maketitle
%

\section{Introduction}

 Relativistic jets from Active Galactic Nuclei (AGN) are among the most powerful and persistent astrophysical outflows \citep[for a review, see][]{Blandford19}. These jets extend from subparsec scales near the supermassive black hole to megaparsec distances, emitting nonthermal radiation across the electromagnetic spectrum and potentially producing high-energy neutrinos and ultra-high-energy cosmic rays \citep[e.g.,][]{SRT09,Petropoulou2020,BM2022}. Despite their remarkable large-scale stability \citep{Willis1974}, jets are susceptible to a broad range of hydrodynamic and magnetohydrodynamic instabilities, including Kelvin–Helmholtz, Rayleigh–Taylor,  centrifugal, current-driven and pressure-driven modes \citep[e.g.][]{Bodo1989, Bodo2004,Bodo2013, Bodo2019, millas2017,Das2019,wang2023,Bodo2021, Musso2024}.
High-resolution observations of nearby radio galaxies, such as M87 and BL Lac, reveal jets that undergo rapid lateral expansion and acceleration near the core, followed by gradual collimation at larger distances \citep{Cohen2014,Walker2018}. This structural evolution is often associated with recollimation shocks, which arise from pressure mismatches between the jet and the surrounding medium \citep{KF1997,BT2018,GK2018,CBT2024,costa25}. These shocks not only influence the jet’s morphology but are also considered prime sites for localized particle acceleration via oblique shocks and stochastic processes in the downstream flow \citep[e.g.,][]{sciaccaluga25}. Observationally, recollimation shocks may also be linked to high-energy variability; for instance, \citet{Hervet2019} proposed that multiple stationary shocks in Mrk 421 could explain complex X-ray flaring patterns, suggesting that recollimation structures actively modulate non-thermal emission. Collectively, these findings underscore the importance of parsec-scale recollimation zones for both the dynamical and radiative properties of AGN jets.
While early models and axisymmetric simulations predicted a stable sequence of recollimation and reflection shocks \citep{KF1997,Mizuno2015,PK2015,2016Fromm}, three-dimensional studies have shown that these structures become unstable beyond the first shock. Simulations reveal the rapid growth of instabilities, the onset of turbulence, and the suppression of multiple shock structures that are commonly observed in 2D \citep{Matsumoto2013,Mizuno2009,GK2018,Matsumoto2021}.
A key mechanism in this context is the centrifugal instability (CFI), triggered by the curvature of streamlines in the recollimation flow. The CFI is intrinsically three-dimensional and can develop even in the absence of a jet–ambient density contrast \citep{GK2018}. Its nonlinear evolution and dependence on jet parameters, however, remain incompletely understood \citep{costa25}.
Magnetic fields further complicate the stability picture \citep{GK2018,Komissarov2019,Gottlieb2020,Matsumoto2021,Hu2024}. In magnetized jets, current and pressure-driven instabilities can emerge and compete with hydrodynamic modes. Recent studies indicate that the kink instability may dominate over conical shock formation in highly magnetized flows, significantly altering jet propagation and energy dissipation patterns \citep{Duran2017}. Linear analyses have provided insights into the growth of these modes under various configurations \citep{Bodo2013, Bodo2019,Kim18,Vlahakis2023,Vlahakis2024, Musso2024}. Overall, the interplay between jet magnetization, velocity shear, density, and the external pressure gradient critically determines jet stability.
In this work, we investigate the onset of instabilities
 in magnetized recollimating flows, under conditions representative of low-power AGN jets, using three-dimensional relativistic hydrodynamic simulations performed with the \texttt{PLUTO} code \citep{PLUTO2007} complemented by a linear stability analysis of the CFI. While previous studies have focused mainly on 2D setups or isolated instabilities, our approach captures the intrinsic 3D character of the CFI, its interaction with magnetic fields, and the transition to turbulence downstream of the first recollimation shock. The linear analysis quantifies the growth rates and parameter dependence of the instability, while the 3D simulations reveal its nonlinear evolution, energy dissipation, and impact on jet structure. By combining these two approaches, we provide new insights into jet stability, the role of magnetic stresses, and the potential connection to observable signatures such as stationary features and high-energy variability.

The paper is organized as follows. In Sect.~\ref{sec:2}, we present the model description and the numerical setup. In Sect.~\ref{sec:2D}, we show the results of our 2D simulations. In Sect. ~\ref{sec:linear} we present the linear analysis of the relativistic CFI. In Sect.~\ref{sec:3D} we present the results of the 3D simulations. In Sect.~\ref{sec:4}, we discuss the implications of the results in the context of jet stability and the development of the CFI and we summarize our conclusions and outline prospects for future work.

\section{Problem description}
\label{sec:2}
Observations suggest the presence of quasi-stationary features in AGN jets on parsec scales, likely associated with recollimation shocks, \citep[see for example][]{Paraschos2025}. These regions have been argued to be the expected location of the high-energy emission zone in blazars. At these distances, variability is typically moderate, occurring on timescales of about a day, which suggests that the emission is produced by a relatively stable dissipation mechanism \citep{Tavecchio2010}. This challenges scenarios involving rapidly propagating shocks and points instead to localized, possibly stationary structures as acceleration sites.

In this work, following the approach adopted in \cite{CBT2024} for a purely hydrodynamical jet, we investigate the recollimation of a relativistic jet at parsec-scale distances from the central engine (the influence of general relativity is negligible in this context). We consider a stage in which the bow shock at the jet head has dissipated and the jet propagates without the protection of a surrounding cocoon. Instead, it interacts directly with the ambient medium, which we interpret as a residual wind or warm interstellar gas still influenced by the gravitational potential of the central source. This environment exerts a confining pressure, for which we assume a decreasing power-law profile, with index $\eta = 0.5$. A free expanding jet is eventually reconfined  by the external pressure, with the formation of a chain of recollimation shocks, \citep{KF1997, PK2015}.

Our methodology is similar in spirit to the works of \citet{GK2018} and \citet{Matsumoto2021}, who also studied the stability of recollimated jets using a two-step simulation process.  In particular, we follow the two-step approach  presented in \cite{CBT2024, costa25}. We note that the equilibrium configuration adopted as the initial state represents an idealized setup that may not be fully realized in nature. This choice, however, allows us to isolate and analyze the growth of specific instabilities under controlled conditions, following the approach of previous works. First, we perform 2D axisymmetric simulations to identify steady-state solutions featuring a clear recollimation structure. 
Specifically, a relativistic conical jet with an opening angle \( \theta_j \) is launched into the computational domain at a distance \( z_0 \) from the cone's apex, embedded within a confining ambient medium. We extend the setup by incorporating magnetic fields into the simulations.
Then, we extend the analysis to 3D, relaxing the symmetry constraints to assess the stability of these solutions and capture the development of instabilities. This strategy allows us to isolate instability growth from jet propagation effects. We note that different procedures for constructing the steady initial configuration, such as starting from an analytical equilibrium or from a numerically relaxed axisymmetric state, may lead to small variations in the early 3D evolution. However, these differences do not affect the qualitative behavior of the instability during its nonlinear development.

The system of equations in conservation form reads
\begin{equation}
\frac{\partial}{\partial t}
\begin{pmatrix}
D \\
\mathbf{m} \\
E_t \\
\mathbf{B}
\end{pmatrix}
+ \nabla \cdot 
\begin{pmatrix}
D\mathbf{u} \\
w_t \Gamma^2 \mathbf{uu} - \mathbf{bb} + \mathbf{I} p_t \\
\mathbf{m} \\
\mathbf{u B} - \mathbf{B u}
\end{pmatrix}
=
\begin{pmatrix}
0 \\
\mathbf{f_g} \\
\mathbf{u \cdot f_g }\\
0
\end{pmatrix}
\end{equation}

We adopt units in which the speed of light $ c = 1 $, and the magnetic field $ \mathbf{B} $ is rescaled to absorb the factor $ \sqrt{4\pi} $. The set of primitive variables is given by the density in the rest frame $ \rho $, the thermal pressure $ p $, the three-velocity $ \mathbf{u} $ in the laboratory frame and the three-vector magnetic field $ \mathbf{B} $ in the laboratory frame.
Here, $ \Gamma $ denotes the Lorentz factor, $ \mathbf{f}_g $ is an external force density in the lab frame, and $ \mathbf{I} $ represents the $ 3 \times 3 $ identity tensor. The conserved variables include the magnetic field $ \mathbf{B} $, as well as the lab-frame mass, momentum, and energy densities, defined as:

\begin{equation}
\begin{aligned}
D &= \Gamma \rho, \\
\mathbf{m} &= w_t \Gamma^2 \mathbf{u} - b_0 \mathbf{b}, \\
E_t &= w_t \Gamma^2 - b_0b_0 - p_t,
\end{aligned}
\qquad \text{with} \qquad
\begin{aligned}
b_0 &= \Gamma\, \mathbf{u} \cdot \mathbf{B}, \\
\mathbf{b} &= \frac{\mathbf{B}}{\Gamma} + \Gamma\, (\mathbf{u} \cdot \mathbf{B})\, \mathbf{u}, \\
w_t &= \rho h + \frac{B^2}{\Gamma^2} + (\mathbf{u} \cdot \mathbf{B})^2, \\
p_t &= p + \frac{\mathbf{B}^2}{2 \Gamma^2} + \frac{(\mathbf{u} \cdot \mathbf{B})^2}{2}.
\end{aligned}
\end{equation}

We close the set of equations with the Taub-Matthews equation of state, which approximates the Synge EoS of a single species relativistic perfect fluid \citep{Mignone2005}: 
\begin{equation}\label{eq:TM}
h=\frac{5}{2}\mathcal{T}+\sqrt{\frac{9}{4}\mathcal{T}^2+1},
\end{equation}
where $h$ is the specific enthalpy and $\mathcal{T}=p/\rho$ is the temperature. 

As explained above at $t=0$ we have a conical jet, in which the velocity is constant and the density and pressure profiles are, following \cite{KF1997}:
\begin{equation}
\rho_{j}(r,z,t=0) = \rho_{j}(0,z_0,0) \left( \frac{R}{R_0} \right)^{-2} ,   
\end{equation}

\begin{equation}
  P_{j}(r,z,t=0) = P_{j}(0,z_0,0) \left( \frac{R}{R_0} \right)^{-2\gamma},  
\end{equation}
where $r$ is the cylindrical radius (measured from the axis of the cone), $R=\sqrt{r^2+z^2}$ is the spherical radius and  $R_0=z_0$. Furthermore, the transverse transition of all the variables from their values in the jet to the values in the external medium are smoothed  to avoid numerical noise at the contact discontinuity, as better specified in Appendix \ref{app:smooth}. The external medium is characterized by density and pressure that
decay with distance from the central object as a power law. The external density profile along $z$ is $\rho_{\text{ext}}(z) =\rho_{ext,0} (z/z_0)^{-\eta}$ and the external pressure profile is $P_{\text{ext}}(z) = P_{ext,0} (z/z_0)^{-\eta}$.

The magnetic field is helical and its components,  in the laboratory frame, are (see \citealt{sciaccaluga25}, for details):
\begin{align}
    &B_r= \frac{\alpha B_0}{R^2}  \text{e}^{-\mathcal{X}^2} \frac{r}{R},\\
    &B_z = \frac{\alpha B_0}{R^2} \text{e}^{-\mathcal{X}^2} \frac{z}{R},\\
    &B_\phi =  \Gamma \frac{B_0}{R}  \sqrt{\text{e}^{-2\mathcal{X}^2} -\frac{\psi_\chi}{2 \sin^2\theta}\left[\psi_\chi-\psi_\chi \text{e}^{-2\mathcal{X}^2}+\sqrt{2\pi}\,\rm{erf}\left(\sqrt{2}\mathcal{X}\right)\right]}
\end{align}
where $\Gamma$ is the Lorentz factor of the flow,  $\mathcal{X} = \left(\cos\theta-1\right)/\psi_\chi$ and $\theta=\cos^{-1}(z/R)$. The parameter $\alpha$ is related to the pitch  and the amplitude $B_0$ is defined in the rest frame, it defines the field strength and is related to the rest frame hot magnetization, at injection, $\sigma$ by the relation
\begin{equation}
    \sigma = \frac{B_0^2}{\rho(z_0) h}.
\end{equation}
Finally, $\psi_\chi$ is the scale length for the decay of the poloidal field, which is assumed to have a Gaussian profile. We set 
$\alpha=1$ meaning that the maximum strengths of the poloidal and toroidal components are comparable in the fluid rest frame, while in the lab frame the toroidal component dominates due to the Lorentz factor.

In our simulations, Tab. \ref{tab:parameters}, we consider a jet with an initial opening angle $\theta_j = 0.1$ and Lorentz factor $\Gamma = 10$. We set $\psi_\chi=10^{-2}\theta_j$, to make the field decay to $0$ outside the jet. The density ratio of the jet to the ambient medium is $\nu = {\rho_j(r=0, z=z_0)}/{\rho_{ext,0}} = 10^{-5}, 10^{-4}$. The pressure ratio is $P_{\text{ratio}} = {P_j(r=0, z=z_0))}/{P_{ext,0}} = 10^{-3}, 10^0, 10^1$ and the magnetization  ranges from 0 (unmagnetized) to 1.

The total jet power $L_{\text{jet}}$ is the sum of the kinetic (matter) power and the magnetic (Poynting) power $L_{\text{jet}} = L_{\text{HD}} + L_B$
where:
$L_{\text{HD}} = \pi z_0^2 \theta_j^2 \rho_j h_j \Gamma_j^2 v_j$ and
$
L_B = \pi z_0^2 \theta_j^2 \frac{B_o^2}{4 \pi} \Gamma_j^2$.
Putting it together, we get:
\begin{equation}
L_{\text{jet}} = \pi z_0^2 \theta_j^2 \Gamma_j^2 \left( \rho_j h_j  v_j + {B_0^2} \right).
\end{equation}
which, for the given parameters, corresponds to low power jets (with a value of $\sim  10^{43} \rm{erg/sec}$, for $z_0=0.1 ~\rm{pc}$ and $\rho_{ext,0}=10^5 \rm{m_p cm}^{-3}$).
\begin{table}[h!]
    \caption{Simulation parameters.}
    \centering
    \begin{tabular}{|c|c|c|c|c|c|}
        \hline
        Case & \boldmath$\nu$ & \boldmath$P_{\mathrm{ratio}}$ & \boldmath$\sigma$ & \boldmath$\Gamma$ & \boldmath$\theta_j$ \\ \hline
        A\_0 & $10^{-5}$ (light) & $10^{-3}$ & $0$ & 10 & 0.1 \\ 
        A\_{0.1} & $10^{-5}$ (light) & $10^{-3}$ & $0.1$ & 10 & 0.1 \\ 
        A\_1 & $10^{-5}$ (light) & $10^{-3}$ & $1$ & 10 & 0.1 \\ \hline
        B\_0 & $10^{-4}$ (heavy) & $10^{-3}$ & $0$ & 10 & 0.1 \\ 
        B\_{0.1} & $10^{-4}$ (heavy) & $10^{-3}$ & $0.1$ & 10 & 0.1 \\ 
        B\_1 & $10^{-4}$ (heavy) & $10^{-3}$ & $1$ & 10 & 0.1 \\ \hline
        C\_0 & $10^{-5}$ (light) & $\sim 1$ & $0$ & 10 & 0.1 \\ 
        C\_{0.1} & $10^{-5}$ (light) & $\sim 1$ & $0.1$ & 10 & 0.1 \\ 
        C\_1 & $10^{-5}$ (light) & $\sim 1$ & $1$ & 10 & 0.1 \\ \hline
        D\_0 & $10^{-5}$ (light) & $10$ & $0$ & 10 & 0.1 \\ 
        D\_{0.1} & $10^{-5}$ (light) & $10$ & $0.1$ & 10 & 0.1 \\ 
        D\_1 & $10^{-5}$ (light) & $10$ & $1$ & 10 & 0.1 \\ \hline
    \end{tabular}
\label{tab:parameters}
    \tablefoot{Each family (A–D) corresponds to a specific combination of density ratio ($\nu$) and pressure ratio ($P_{\mathrm{ratio}}$), while the suffix indicates the magnetization $\sigma$.}
\end{table}

The numerical simulations are carried out by solving the equations of relativistic ideal magnetohydrodynamics first in 2D and then in 3D, by using the ideal RMHD module of the PLUTO code \citep{PLUTO2007}.  The equations were evolved with second-order Runge-Kutta time stepping, linear reconstruction, and the HLLD Riemann solver \citep{Mignone2009}. Beyond these methods, magnetic field evolution required additional care. Constrained transport \citep{Balsara1999, Londrillo2004} was used to maintain the condition $\nabla \cdot \mathbf{B} = 0$ 

The setup of each 2D simulation uses cylindrical coordinates $(r, z)$, with a domain 
$[0, L_{r,2D}] \times [1, L_{z,2D}]$. All lengths are expressed in units of $z_0$, 
defined as the distance from the cone (Appendix \ref{app:3dres} for details). 
 At the final time, the 2D simulations provide a nearly steady state which is used as the  initial state for the 3D simulations. 3D simulations were carried out on a Cartesian domain with coordinates in the range $x \in [-L_{x,3D}/2, L_{x,3D}/2]$, $y \in [-L_{y,3D}/2, L_{y,3D}/2]$, and $z \in [1, L_{z,3D}]$ (lengths are expressed in units of the $z_0$). Here $L_x = L_y$, while we chose $z$ as the jet propagation direction. We employed a grid of $N_x \times N_y \times N_z$ zones with a uniform grid spacing along the $z$ (axial) direction (Appendix \ref{app:3dres} for details).

\section{Results}
To build a coherent picture of jet stability, we start from the simplest setting and add complexity step by step. Our analysis begins with 2D steady-state solutions, which show that even jets with the same Lorentz factor, opening angle, and magnetization can develop different internal structures depending on the density and pressure ratios.  Therefore, they may also have different stability properties with respect to the centrifugal instability (CFI).  
We then derive a local onset criterion and perform a linear stability analysis to determine how the growth rate of the CFI depends on jet parameters. With this insight, we then select two contrasting 3D simulations—one prone to disruption and one remaining stable—allowing us to investigate how magnetization, curvature, and external confinement jointly govern jet stability.

\subsection{Insights from the 2D simulations}\label{sec:2D}
In this section, we present the main findings from our two-dimensional relativistic MHD simulations, which are used to construct steady jet solutions under different physical conditions. By varying the density ratio, pressure, and magnetization, we identify equilibria that serve as a baseline for assessing stability in three dimensions and for comparison with linear analysis. At the same time, the 2D runs already provide valuable insight into how the centrifugal instability (CFI) can be triggered or suppressed, highlighting parameter combinations that lead either to long-lived stable flows or to configurations prone to disruption.
Although two-dimensional simulations cannot capture the full spectrum of three-dimensional modes, they reproduce many of the essential dynamical features of relativistic jets. These include the formation of recollimation shocks, oscillatory jet profiles, and the redistribution of magnetic and thermal pressure. In particular, the 2D solutions offer a  diagnostic of where strong curvature and flow gradients develop  -- precisely the conditions that are expected to enhance the growth of centrifugal and related instabilities.
\begin{figure}[htbp]
\resizebox{\hsize}{!}{\includegraphics[clip=true]{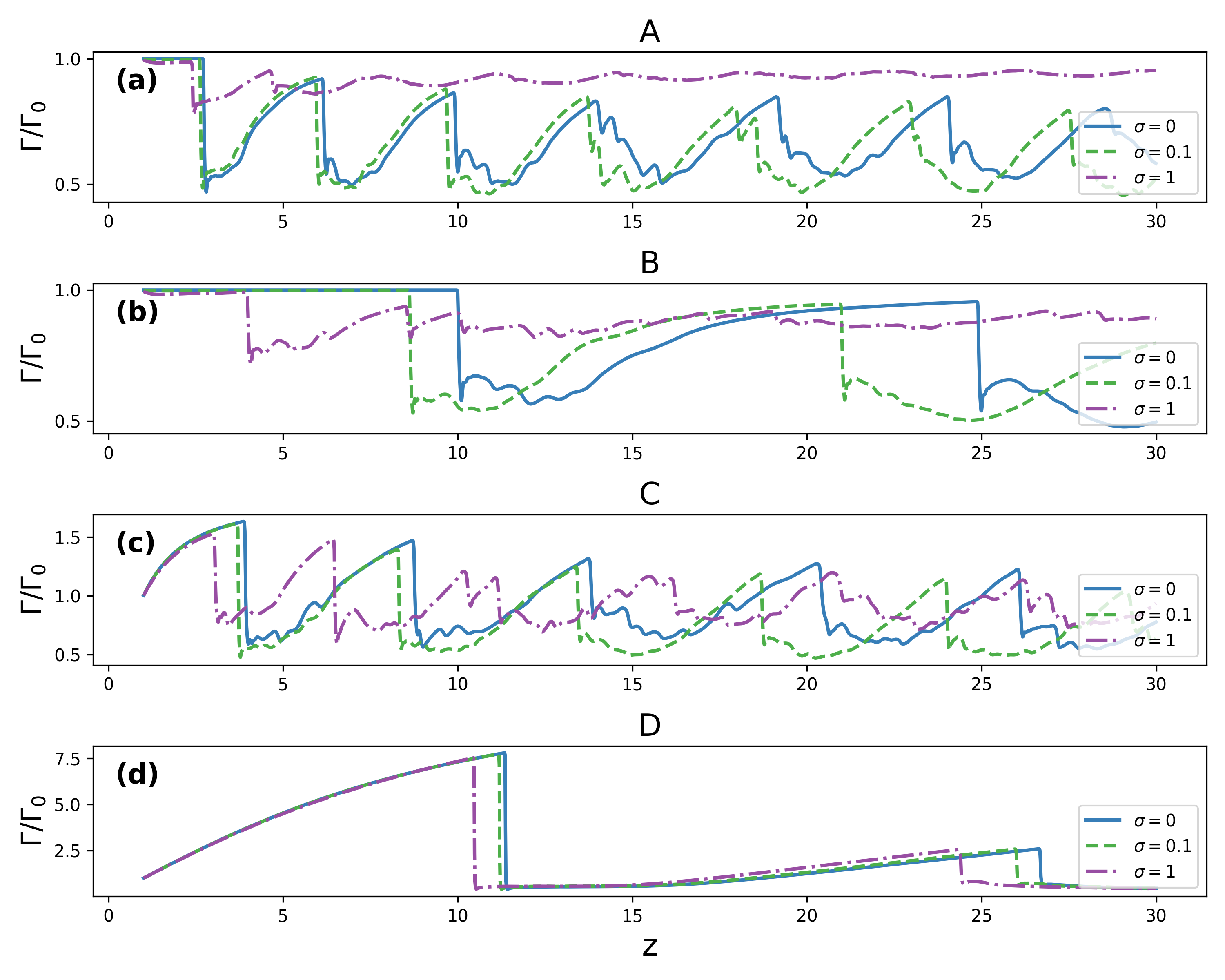}}
\caption{\footnotesize{Profiles along the $z$-axis at position $x=0$ of the normalized Lorentz factor for the 2D simulations presented in Tab.\ref{tab:parameters}. The normalization is taken at the position $(0,0)$ for simulations A, B, C, and D with magnetization $\sigma = 0$, $0.1$, and $1$.}}
\label{fig:2Dcomp}
\end{figure}

Figure~\ref{fig:2Dcomp} shows the normalized Lorentz factor profiles along the $z$-axis at $x=0,  y=0$ for simulations A, B, C, and D. The normalization is taken at the injection point $(0,0)$. Despite identical initial Lorentz factors and jet opening angles within each set, the location of the first recollimation shock shifts noticeably between cases. This shift is solely due to changes in the density ratio, temperature and magnetization, indicating that these parameters influence the effective curvature radius and radial expansion rate of the jet. 

In cases A (cold, light case) the first recollimation point is the closest to the origin. We also observe that  the Lorentz factor shows a steady decrease on top of the oscillations due to recollimation shocks. 
A similar evolution occurs in cases B: the first recollimation point is somewhat shifted while the larger inertia moderates the global  deceleration relative to cases A.
Cases C is warm ($kT \sim m_e c^2$) and injected close to pressure equilibrium with the ambient and we observe a slight radial expansion 
that leads to a conversion of thermal to kinetic energy and to an increase of the Lorentz factor.

Once the jet recollimates, however, a fraction of this kinetic energy is reconverted into internal energy and the Lorentz factor shows a global decrease that is even larger than in cases A. In cases D the jet is also warm but injected with a pronounced overpressure, that drives a significant radial expansion and efficient acceleration:  
the Lorentz factor can reach $\Gamma \simeq 75$ before the recollimation point. 
In this case, the thermal pressure plays a crucial role in converting internal energy into bulk motion.

Across all cases, a common trend emerges: despite the different injection conditions, the flow is not able to reconvert all the thermal energy into kinetic form. Increasing the magnetization strengthens the role of magnetic tension, which provides 
a more effective confinement of the jet and shifts the recollimation shocks to smaller axial distances; this behaviour is evident in all the cases shown in Fig.~\ref{fig:2Dcomp}. A backward displacement of the shock is visible in each case, although the effect is 
very small in cases~A. Since it appears systematically, it is reasonable to attribute it to the same underlying mechanism, likely related to the tension of the toroidal magnetic component. The detailed dependence of the shock location on magnetization will be presented in a forthcoming paper (Boula et al., in prep.).

\begin{figure*}
\centering
\includegraphics[width=\linewidth,clip=true]{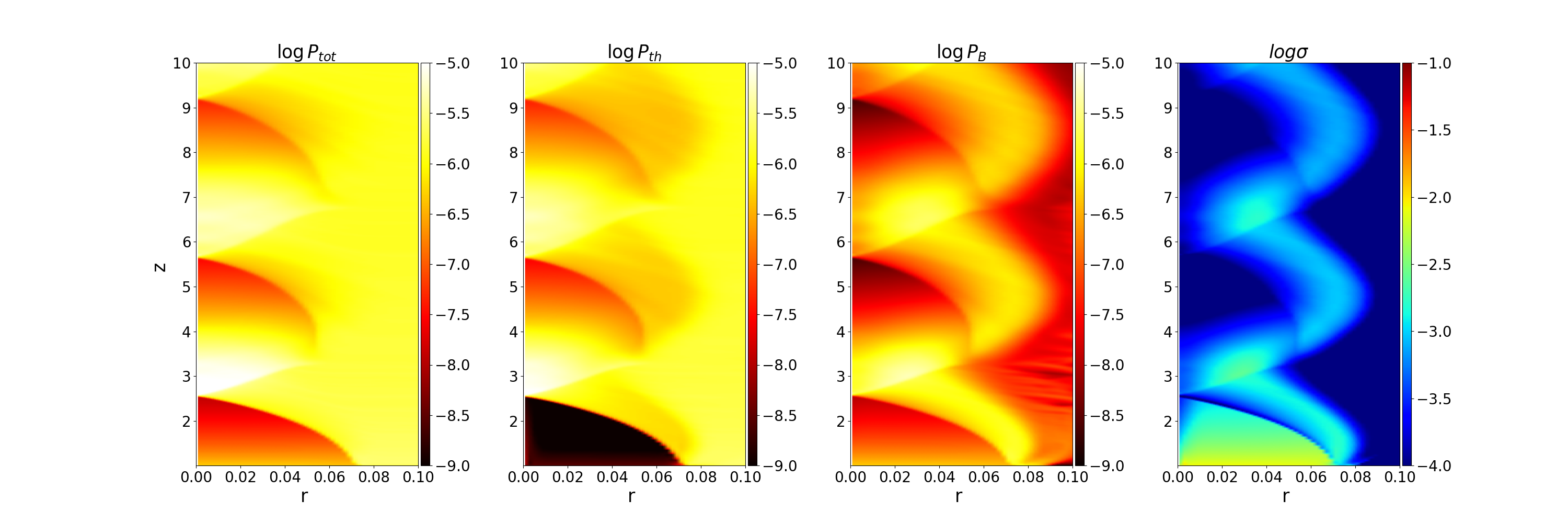}
\caption{For the 2D simulation A\_0.1 with magnetization $\sigma = 0.1$, the logarithmic maps of total pressure, thermal pressure, magnetic pressure, and total magnetization are shown. The initial parameters are: density ratio $\nu = 10^{-5}$, Lorentz factor $\Gamma = 10$, jet opening angle $\theta_\text{jet} = 0.1$, and pressure ratio $P_\text{ratio} = 10^{-3}$. The maps correspond to the final time step $t = 3000 \, t_\text{cr}$ in units of $z_0/c$.}
\label{fig:rec.point}
\end{figure*}

The internal structure of a representative case is shown in Fig.~\ref{fig:rec.point} for simulation A\_0.1 with $\sigma = 0.1$. We adopt $\sigma = 0.1$, as this choice leads to strong recollimation shocks, unlike the case with $\sigma = 1$. The logarithmic maps of total, thermal, and magnetic pressure, along with the total magnetization, reveal how different pressure components dominate in different regions. Upstream of the recollimation shock, thermal pressure contributes to acceleration as the flow  expand, while magnetic pressure becomes more significant downstream of the shock, particularly along the jet edges. At the recollimation shock, magnetic pressure enhancements appear at the boundary layers, plausibly related to compression and flow deflection. A contribution from magnetic tension cannot be excluded, and this aspect will also be investigated in more detail in the forthcoming paper (Boula et al., in prep.).

\subsection{Centrifugal Instability (CFI)}\label{sec:linear}

This study focuses on the centrifugal instability (CFI), a local instability that develops in curved flows. The presence of a toroidal magnetic field may, however, lead to stabilization due to the competing effect of magnetic tension.

To quantify the local conditions for CFI, we perform an order-of-magnitude comparison between the restoring magnetic tension due to the toroidal component of the magnetic field and the effective centrifugal force due to the motion along the curved streamlines. 
In a cylindrical or weakly perturbed geometry, the magnetic tension per unit volume scales as 
$B_{{\phi}}^2/R_j$, while the inertial term associated with curved motion scales as
$(\Gamma^{2}\rho_j h_j+B^2)/R_c$, where $B$ is the total magnetic field and $B_{{\phi}}$ is its toroidal component 
$R_j$ is the local jet radius, and $R_c$ the streamline curvature radius \citep{Komissarov2019, Matsumoto2021}. 
Equating these contributions yields:
\begin{equation}
 \frac{B_{\phi}^2}{R_j} \sim \frac{(\Gamma^2\rho h +B^2)}{R_c}
    \rightarrow
    \frac{\sigma_{\mathrm{tor}}}{(\sigma+1) R_j} \sim \frac{\Gamma^2}{R_c}
    \rightarrow
    \frac{\sigma_{\mathrm{tor}}}{\Gamma^2(\sigma+1)} \sim \frac{R_j}{R_c},
\end{equation}
where $\sigma$ and $\sigma_{tor}$ are respectively the local values of the full magnetization and of the magnetization related to the toroidal component of the field.
This shows that $\sigma_{\mathrm{tor}}/\Gamma^2$ serves as a direct measure of the competition 
between stabilizing magnetic tension and destabilizing curvature (the total magnetization $\sigma$ is less than the unity, and we can be neglected its contribution in the denominator).

\begin{figure}[htbp]
\resizebox{\hsize}{!}{\includegraphics[clip=true]{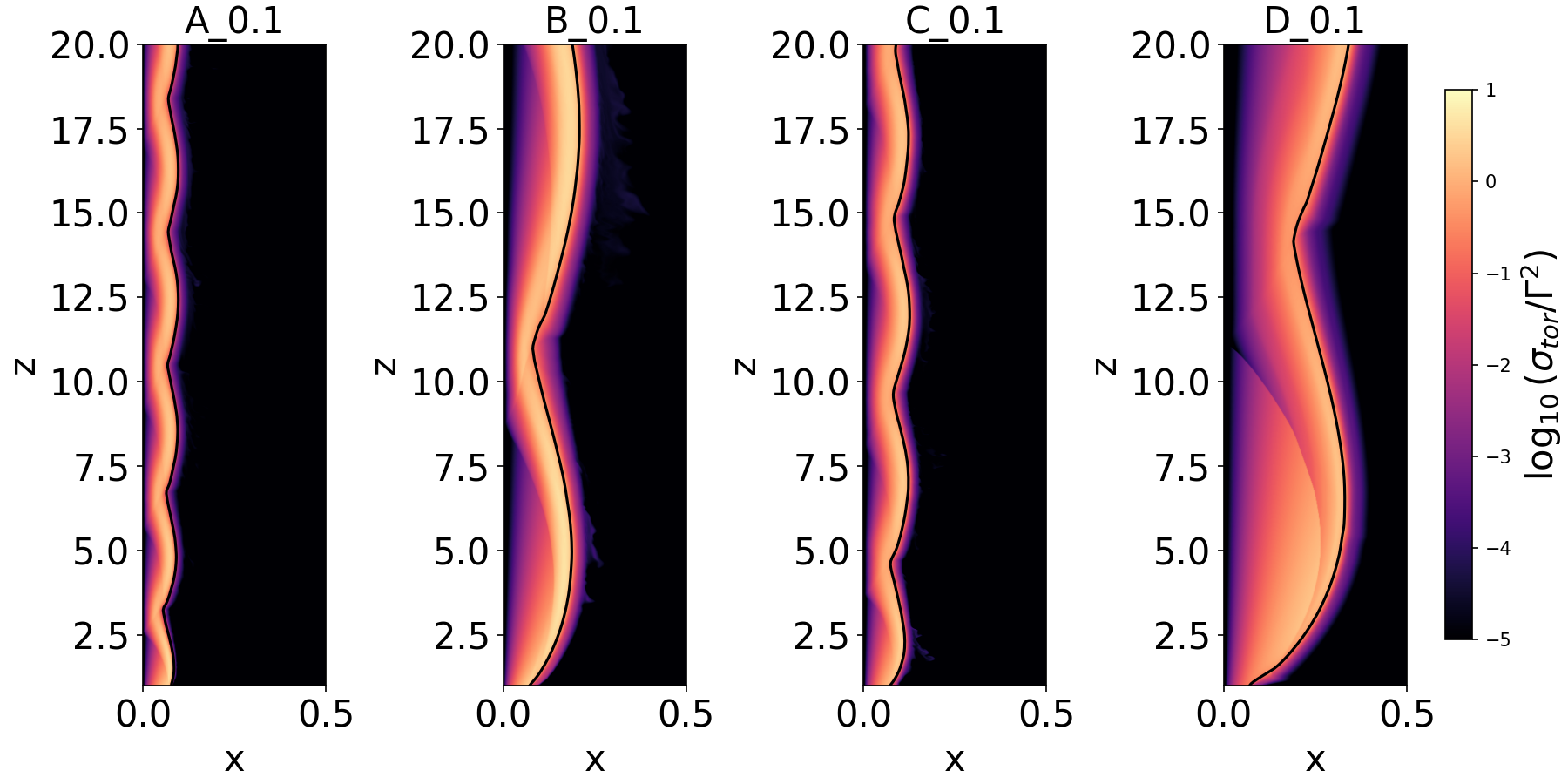}}
\caption{\footnotesize{Streamlines (initialized from the same starting point) overlaid on the spatial distribution of $\sigma_{\mathrm{tor}}/\Gamma^2$ for all simulation runs. This ratio serves as a local diagnostic for the onset of centrifugal instability, with low values marking regions where the jet is most susceptible to curvature-driven perturbations. In the widest jets (e.g. Case~D\_0.1), outer streamlines would likely reach regions satisfying the CFI criterion; here we show a representative streamline to illustrate the overall trend.}}
\label{fig:2Dsigma01}
\end{figure}

A complementary perspective is provided in Fig.~\ref{fig:2Dsigma01}, which shows streamlines—initialized from the same starting point—overlaid on the spatial distribution of $\sigma_{\mathrm{tor}}/\Gamma^2$ for each simulation. Using identical seeds ensures a direct visual comparison of streamline curvature and magnetization across cases. However, for the quantitative analysis presented below, we vary the initial streamline position to ensure that the selected trajectories actually pass through the regions favorable to the development of the CFI. For clarity, the figure highlights a representative streamline that illustrates the systematic trend across simulations.

The quantity $\sigma_{\mathrm{tor}}/\Gamma^2$ quantifies the relative importance of toroidal magnetic tension to inertia and serves as a local diagnostic for the onset of the CFI. Regions with large $\sigma_{\mathrm{tor}}/\Gamma^2$ are more resistant to bending and tend to suppress CFI growth. In contrast, regions of strong curvature and velocity shear often correspond to lower values of $\sigma_{\mathrm{tor}}/\Gamma^2$, which makes them more favorable for the onset of the instability. In several runs, such regions appear already upstream, close to the local minimum of the jet radius, where perturbations can be seeded. In fully three-dimensional flows, these perturbations may be further amplified by Richtmyer–Meshkov–type interactions as the jet passes through the reflection shock close to the the minimum radius, \citep{Matsumoto2013}.

In table \ref{tab:sigma_gamma_radius} we report, for each run, the normalized magnetization $\sigma_{\mathrm{tor}}/\Gamma^2$, the maximum transverse displacement $R_{\text{j,max}}$, the local curvature radius $R_{\text{c}}$ (computed via circle fitting using ten points centered on $R_{\text{j,max}}$), and their ratio $R_{\text{j,max}}/R_{\text{c}}$. The last column lists the dimensionless combination $(\sigma_{\mathrm{tor}}/\Gamma^2)/(R_{\text{j,max}}/R_{\text{c}})$, which we use as a proxy for the balance between magnetic tension (stabilizing) and streamline curvature (destabilizing).

\begin{figure}[htbp]
\begin{center}
\includegraphics[width=\linewidth]{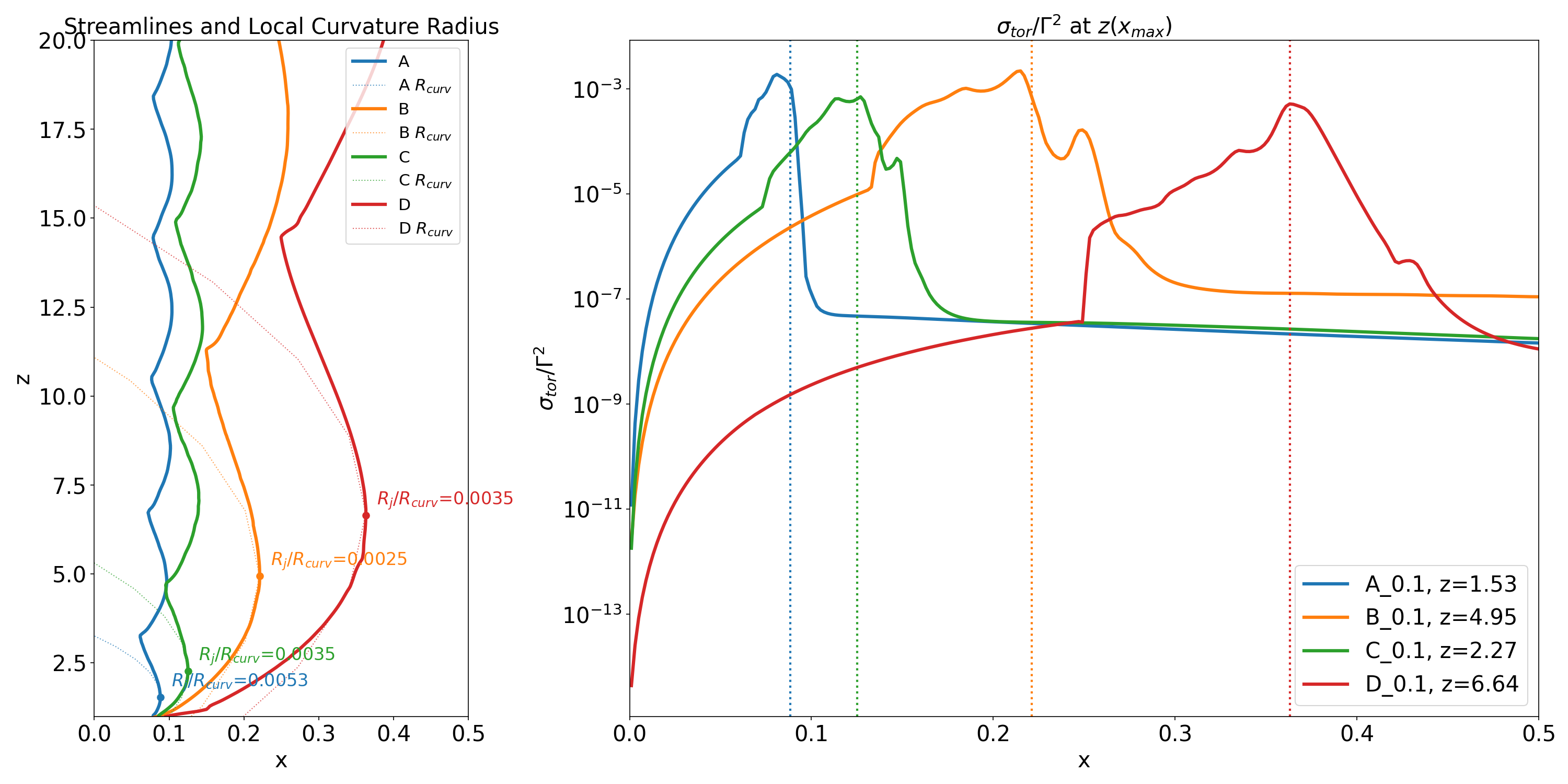}
\caption{
{Left:} Streamlines for cases A\_0.1--D\_0.1 (as in Fig.~\ref{fig:2Dsigma01}), starting at 
$x_0 = 0.082, 0.098, 0.091,$ and $0.124$ respectively, with curvature 
and the ratio $R_{\mathrm{j,max}}/R_{\mathrm{c}}$. 
{Right:} Profiles of $\sigma_{\mathrm{tor}}/\Gamma^{2}$ versus $x$ 
at $z = z_{x_{\max}}$, with vertical lines marking 
$x_{\max} = R_{\mathrm{j,max}}$.
}
\label{fig:compcfi}
\end{center}
\end{figure}

All quantities are evaluated at the axial location of maximum jet displacement (marked by the vertical lines in Fig.~\ref{fig:compcfi}). This choice ensures consistency across runs, and test calculations indicate that this region is the most favorable for the development of the instability. The magnetization $\sigma_{\mathrm{tor}}$ is measured at the jet boundary along this axial position, where curvature effects are strongest.

\begin{table}[h!]
\caption{Numerical results for the CFI condition.}
\centering
\begin{tabular}{c|c|c|c|c|c}
Case & $\boldsymbol{\sigma_{\mathrm{tor}}/\Gamma^2}$ & $\boldsymbol{R_{\text{j,max}}}$ & $\boldsymbol{R_{\text{c}}}$ & $\boldsymbol{R_{\text{j,max}}/R_{\text{c}}}$ & Instability proxy\\
 & & & & & $\dfrac{(\sigma_{\mathrm{tor}}/\Gamma^2)}{(R_{\text{j,max}}/R_{\text{c}})}$\\
A\_0.1 & 0.00112  & 0.09  & 16.7 & 0.0053 & 0.21 \\
B\_0.1 & 0.00073  & 0.22  & 87.1 & 0.0025 & 0.29 \\
C\_0.1 & 0.00065  & 0.13  & 36.2 & 0.0035 & 0.19 \\
D\_0.1 & 0.00062 & 0.36  & 103.0 & 0.0035 & 0.15 \\
\end{tabular}
\tablefoot{Comparison of the normalized magnetization $\sigma_{\mathrm{tor}}/\Gamma^2$ and curvature-based geometric quantities across simulation runs, Fig. \ref{fig:compcfi}. The last column shows the ratio $(\sigma_{\mathrm{tor}}/\Gamma^2)/(R_{\text{j,max}}/R_{\text{c}})$, used as a qualitative proxy for centrifugal instability strength.}
\label{tab:sigma_gamma_radius}
\end{table}

\begin{figure}[htbp]
\centering
\includegraphics[width=\linewidth]{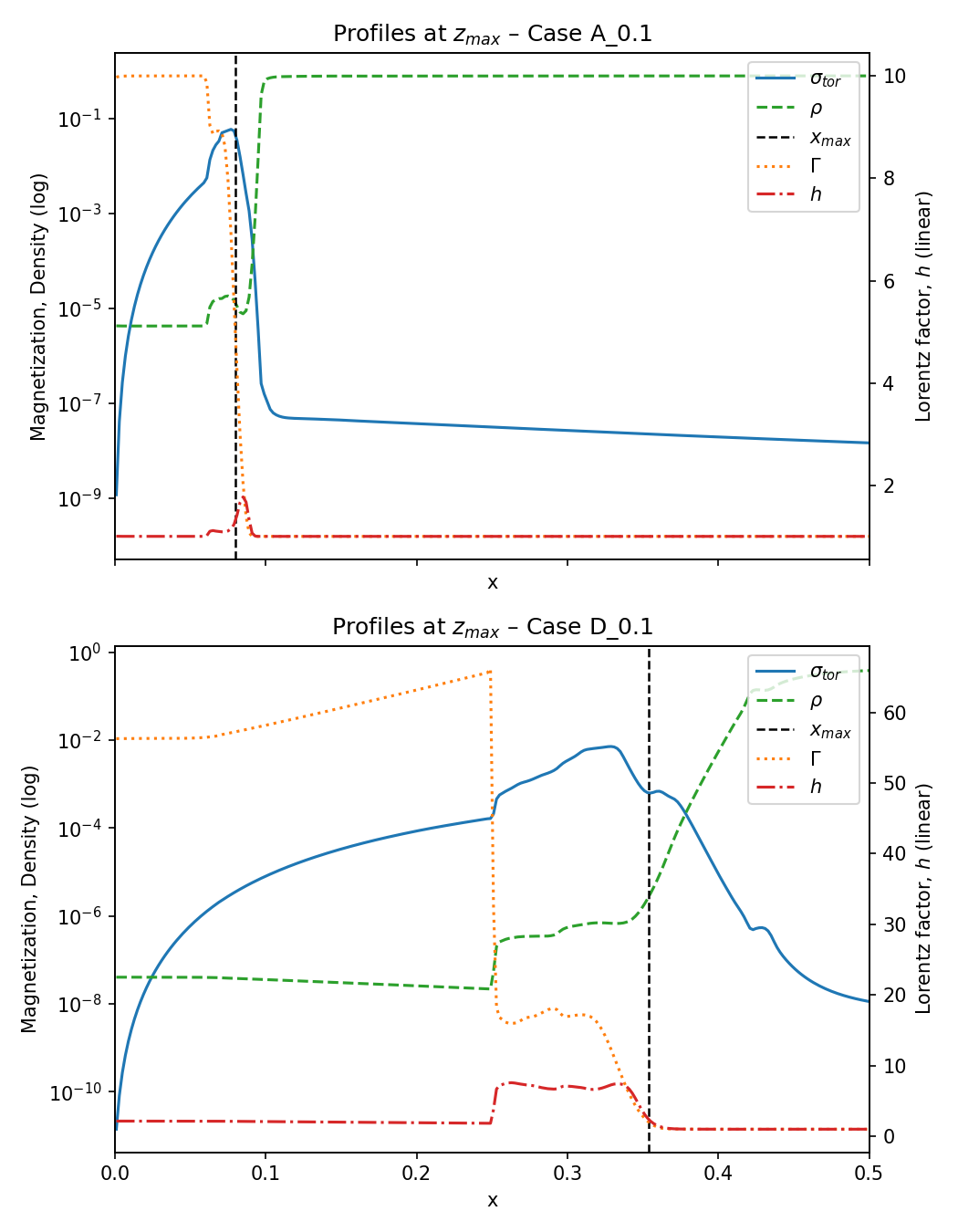}
\caption{Comparison of radial profiles of toroidal magnetization 
$\sigma_{\mathrm{tor}}$ (blue), Lorentz factor $\Gamma$ (orange), 
rest-mass density $\rho$ (green), and enthalpy $h/\rho$ (red) 
at $z = z_{\max}$ for Case~A\_0.1 (top) and Case~D\_0.1 (bottom). 
Case~A\_0.1 shows smooth, monotonic profiles, whereas Case~D\_0.1 exhibits 
sharp gradients and non-monotonic behavior, underscoring the 
complexity of its internal structure. The vertical dashed line 
marks the position of $x_{\max}$ determined from the streamline analysis. 
Profiles are plotted on a logarithmic scale.}
\label{fig:2DD}
\end{figure}

Table~\ref{tab:sigma_gamma_radius} indicates that all four cases are unstable, with the instability proxy reflecting the combined effects of $\sigma_{\mathrm{tor}}/\Gamma^2$ and curvature radius. Figure~\ref{fig:2DD} compares the radial profiles of $\sigma_{\mathrm{tor}}$, $\Gamma$, $\rho$, and $h$ at $z = z_{\max}$ for Cases~A\_0.1 and D\_0.1. Both cases exhibit smooth, but Case~D\_0.1 displays a wider and more complex radial transition region outside the recollimation shock. This illustrates that even when the instability proxy is similar, the spatial structure of the jet can vary, which may influence the growth and localization of local instabilities.

According to the \cite{Komissarov2019} criterion, suppression of CFI modes with azimuthal wavenumber $m \ge 4$ at the interface between reconfined jets and external gas occurs for $\sigma_{\mathrm{tor}}/\Gamma^2 \lesssim \theta_0^2/16$, where $\theta_0$ is the initial half-opening angle of the jet. In our simulations $\theta_0^2/16 \simeq 0.000625$. Comparing this value with  $\sigma_{\mathrm{tor}}/\Gamma^2$ (see Tab.\ref{tab:sigma_gamma_radius})
all cases are stable according to the \cite{Komissarov2019} criterion. This parametrizes the effect of geometry on the instability by using $\theta_0$, but our results highlight the importance of the local geometry (i.e. the local curvature and jet radius) when studying the stability. In turn, this is determined not only by $\theta_0$ but also by the dynamics of the jet.

To further quantify these findings, we now turn to a linear stability analysis aimed at computing the CFI growth rate as a function of jet parameters. This approach allows us to connect the local diagnostics discussed above with the expected timescale for instability development.

\subsection{Linear analysis}

Similarly to \cite{Komissarov2019}, we examine the stability of a magnetized rotating cylindrical shell with a purely axial magnetic field. In contrast to that work, we carry out a linear stability analysis to compute the growth rate of the centrifugal instability, following the methodology of \cite{Vlahakis2023}\footnote{The procedure follows the steps outlined in Sect. 3 of \cite{SV2023}. The only difference is that the integration here starts from $r = 1 - \Delta r$ with $y_1 = 0$, and the outer Hankel functions are replaced by modified Bessel functions because $\lambda$ is purely imaginary in our case. This change is minor and is handled automatically in the computation.}.
\begin{figure}[htbp]
\begin{center}
   \includegraphics[width=\linewidth]{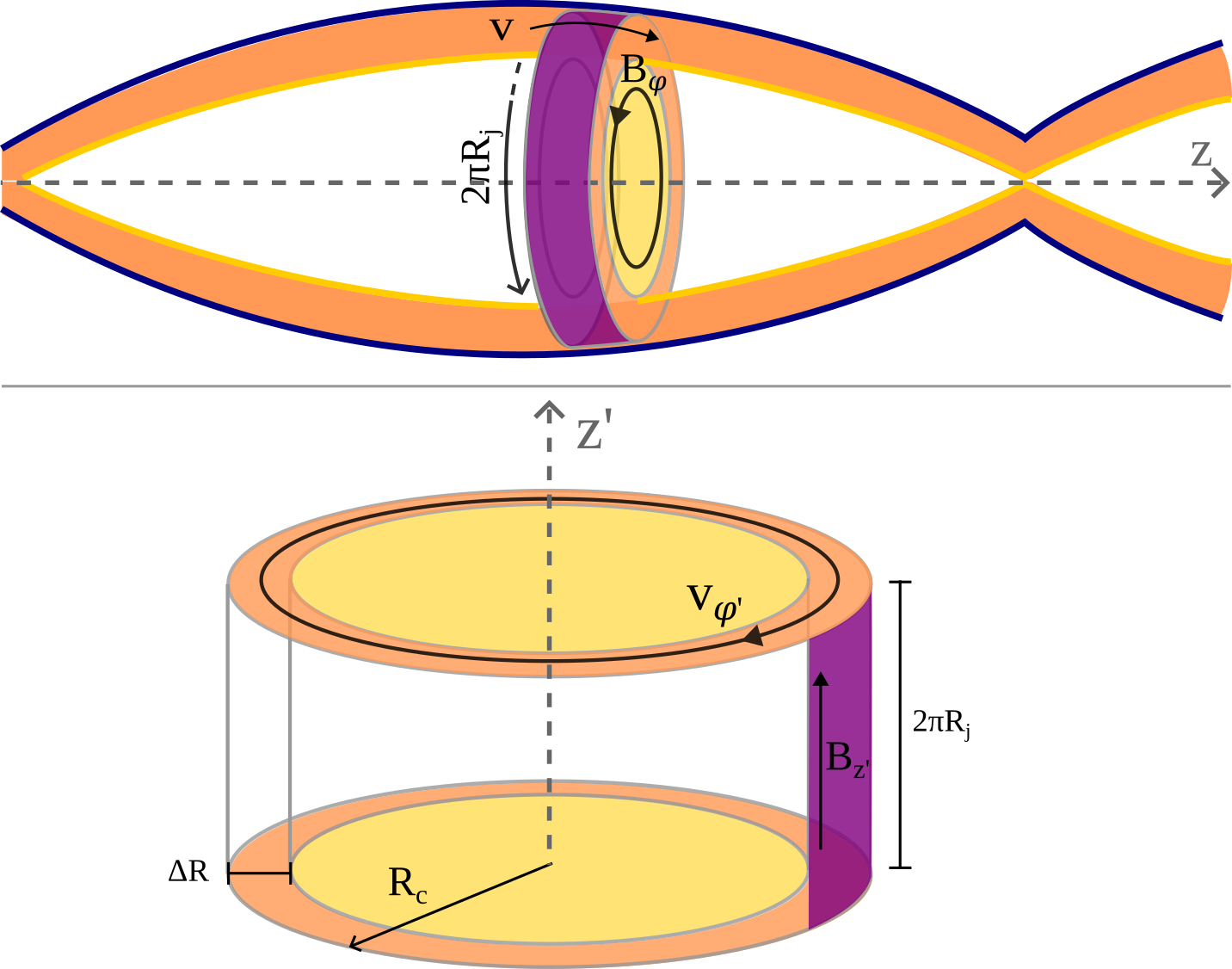}
\caption{Cartoon illustrating the configurations used in the linear analysis, adapted from \cite{Komissarov2019}. 
  {Top panel:} A relativistic jet with a purely azimuthal magnetic field undergoes recollimation due to the confining action of an external pressure. The white inner region represents the freely expanding, unshocked flow, while the shaded area corresponds to the shocked outer layer. The interface between the two is the conical reconfinement shock driven into the jet. 
 {Bottom panel:} A magnetised, rotating cylindrical shell with a purely axial magnetic field, representing the idealised configuration analysed in this work.\label{fig:CFI_c}}
\end{center}
\end{figure}

We adopt a cylindrical coordinate system $(R, z)$ centered on the shell axis, distinct from the jet frame used elsewhere in this paper (Fig. \ref{fig:CFI_c} for details). The shell has outer radius $R_{\text{c}}$, width $\Delta R$, and rotates with a constant azimuthal velocity $V_\phi$, corresponding to a Lorentz factor $\Gamma$. Its specific enthalpy $h$ and magnetization $\sigma$ are taken to be uniform. The magnetic field is expressed as $\mathbf{b}=\sqrt{\sigma h\rho}\hat{z}$, and the thermal pressure as $P = \rho \mathcal{T}$. Radial force balance then yields a power-law density profile $\rho(R) = \rho_j (R/R_{\text{c}})^a$, with exponent
$a={(\Gamma^2-1)(1+\sigma)}/({\mathcal{T}/h+\sigma/2})$
We fix the shell width as $\Delta R = R_{\text{c}}/a$.

The shell is embedded in an external medium characterized by constant density $\rho_{\text{ext}}$, specific enthalpy $h_{\text{ext}}$, and a uniform axial magnetic field with strength $\sqrt{\sigma_{\text{ext}} h_{\text{ext}} \rho_{\text{ext}}}$. Pressure equilibrium at the shell boundary gives the density ratio:
$
{\rho_{\text{ext}}}/{\rho_{R_c}} = ({\mathcal{T}_{R_c} + \sigma_{R_c} h_{R_c}/2})/({\mathcal{T}_{\text{ext}} + \sigma_{\text{ext}} h_{\text{ext}}/2}).
$
This ratio, and in particular the quantity $h_e \equiv h_{\text{ext}}/h$, controls the density contrast between the shell and its surroundings. For minor deviations from $h_e \approx 1$, the environment has little effect; more substantial effects appear only when $h_e \gg 1$.

Since the CFI eigenfunction is localized near the shell’s outer edge, we model the inner boundary as a rigid wall. Perturbations of the form $f(R) e^{i(kz - \omega t)}$ are introduced in the linearized equations, which are then solved across the shell and the external region using appropriate matching conditions, as described in \cite{Vlahakis2023}. The resulting eigenvalue problem yields the (purely imaginary) instability growth rate $\Im \omega$ as a function of the wavenumber $k$.

\begin{figure}[htbp]
\begin{center}
   \includegraphics[width=\linewidth]{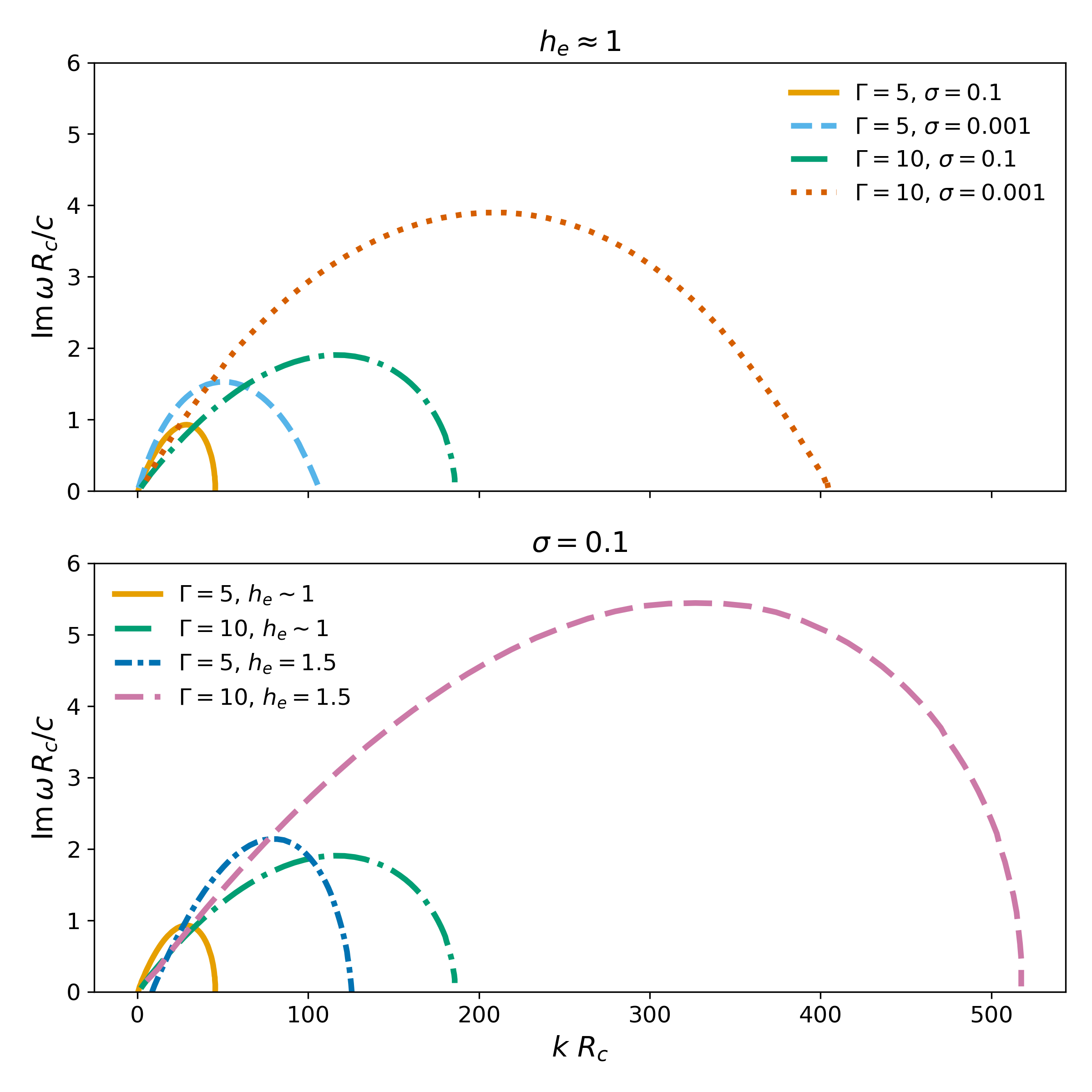}
   \caption{Results from the linear analysis of the centrifugal instability (CFI). {Top panel:} Growth rate as a function of $k ~R_c$, for varying Lorentz factors and magnetizations, assuming fixed external conditions ($h_e \approx 1$). {Bottom panel:} Same as top, but for fixed magnetization $\sigma = 0.1$, and varying external enthalpy ratio $h_e$ and Lorentz factor $\Gamma$. \label{fig:linear}}
\end{center}
\end{figure}

The results are presented in Fig.~\ref{fig:linear}. The top panel displays the growth rate for various values of $\Gamma$ and $\sigma$, with the density contrast held constant ($h_e \approx 1$). As shown, increasing the Lorentz factor $\Gamma$ or decreasing the magnetization $\sigma$ significantly enhances the instability, as both the peak growth rate and the unstable wavenumber range increase. This behavior is consistent with the qualitative criterion proposed by \cite{Komissarov2019}, in which the growth of the CFI is linked to the ratio $(\sigma_{\mathrm{tor}}/\sigma+1)/\Gamma^2 \sim \sigma_{\mathrm{tor}}/\Gamma^2$, since $\sigma< 1$.

The bottom panel illustrates the effect of the external medium by fixing $\sigma = 0.1$ and varying both $\Gamma$ and the enthalpy ratio $h_e$. A comparison between $h_e \approx 1$ and $h_e = 1.5$ shows that the environment has a negligible impact unless $h_e$ exceeds unity. For example, the growth rate curves for $h_e \approx 1~(h_e -1 =10^{-4})$ are indistinguishable, while the case $h_e = 1.5$ exhibits a marked increase in both growth rate and unstable range—especially for higher Lorentz factors. The choice of $h_e = 1.5$ reflects conditions in the simulations, where the external enthalpy is measured to reach values around 1–1.5; in practice, this translates to a corresponding contrast in density between jet and environment.

Overall, these results confirm that centrifugal instability is most effective in fast, weakly magnetized flows and is largely insensitive to the external medium, unless the density contrast is large. This is consistent with the parameter ranges used in our linear analysis, which were motivated by simulation values: $\Gamma = 5$–10, $\sigma = 0.001$ (almost hydrodynamic) or 0.1 (moderately magnetized), and external enthalpy $h_e$ characteristic of 1–1.5.

\subsection{3D results}\label{sec:3D}
\subsubsection{The case run A\_0.1}\label{sec:3DA}

As a reference, we adopt a moderate magnetization $\sigma = 0.1$, which lies between the nearly hydrodynamic regime (small $\sigma$) and strongly magnetized flows prone to kink instabilities \citep[e.g.,][]{Duran2017}. In this regime, we  focus on CFI instabilities, which  should be stable according, for example, to \citet{Matsumoto2021}. As already discussed, we use as initial condition for the 3D simulations the steady state reached at the end of the 2D simulations that we present (for case A\_0.1) in Figure \ref{fig:rec.point}, where we  display maps of the total pressure, thermal pressure, magnetic pressure, and total magnetization. These plots focus on the region near the base of the jet and the first three recollimation shocks. We observe that the recollimation shock is strong and produces a distinct radial structure in the jet. At the jet edges, the total pressure is dominated by magnetic pressure, while along the axis, thermal pressure is stronger. 
\begin{figure}[htbp]
\begin{center}  \includegraphics[width=\linewidth]{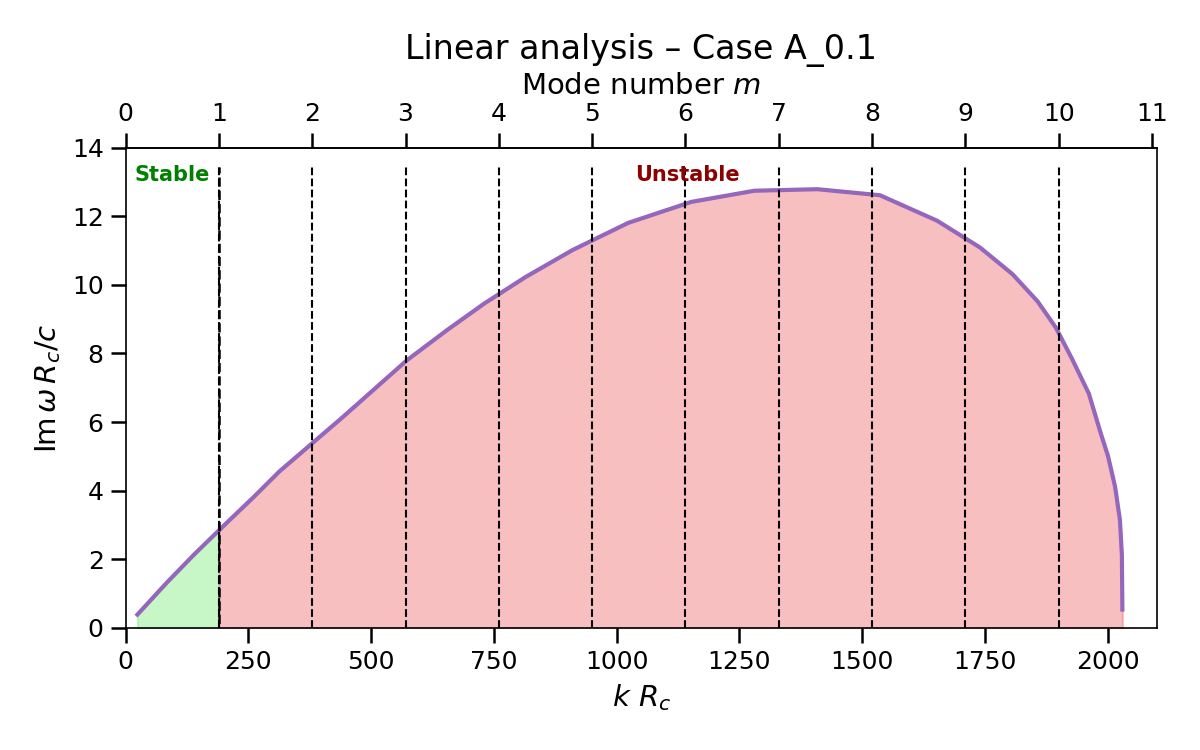}
   \caption{Growth rate of the centrifugal instability (CFI) as a function of axial wavenumber $k ~R_c$, computed using the profiles for simulation A\_0.1 presented in Fig.~\ref{fig:2DD}. Shaded vertical bands indicate the values of $k ~R_c$ corresponding to azimuthal mode numbers $m = 1$ to $m > 6$, derived from the simulation jet radius $R_j = 0.08$ and local curvature radius $R_c = 15.28$. The most unstable mode occurs at $k_{\text{max}} ~R_c \approx 190$, in agreement with the effective $k ~R_c $ from the simulation for $m = 1$. 
   \label{fig:A_linear}}
\end{center}
\end{figure}

\begin{figure}[htbp]
\resizebox{\hsize}{!}{\includegraphics[width=\linewidth]{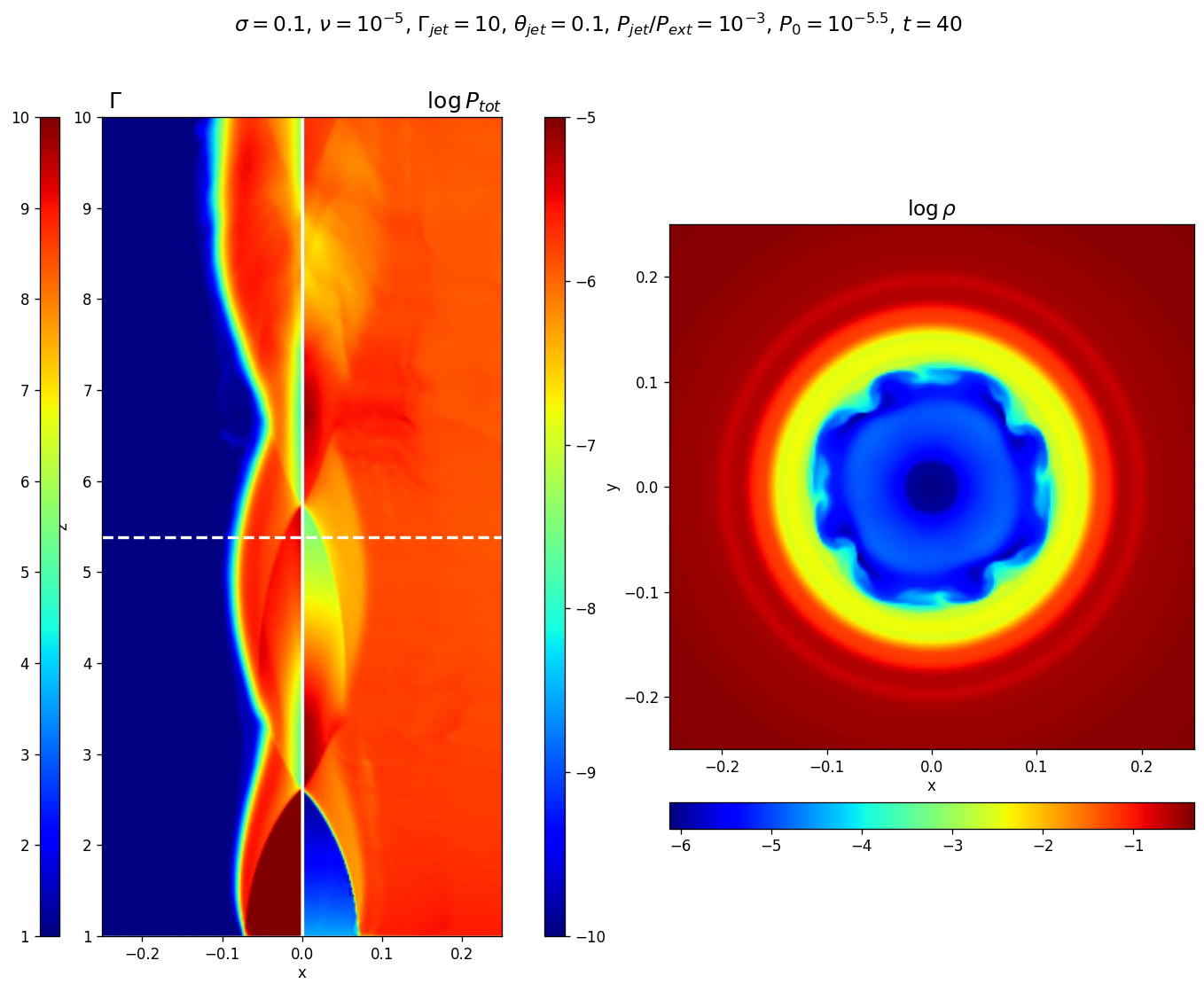}}
\caption{{For the 3D case A\_0.1 with magnetization $\sigma = 0.1$, the first column shows maps of $\Gamma$ and $\log P_\text{tot}$, with transverse cuts of $P_{th}$ and $\log \rho$ at a $z$-position. The horizontal line corresponds to these cut, showing the logarithmic density.}}
\label{fig:3Dmapsprofs}
\end{figure}
The pressure maps in Figure~\ref{fig:rec.point} confirm that during the early free expansion phase, magnetic pressure dominates throughout most of the jet. After recollimation, thermal pressure becomes  important, particularly near the axis.

We can perform a linear stability analysis for parameters extracted from the initial jet configuration. In the linear analysis, the instability is studied in the idealized configuration of a rotating cylindrical shell of height $2\pi R_j$, so that perturbations can be decomposed into Fourier modes with axial wavenumber $k$, Fig. \ref{fig:CFI_c}, bottom panel. In our simulations, however, we probe the instability through azimuthal structures developing around the jet cross-section, which requires translating the axial description of the linear model into an azimuthal one. We identify the cylinder in the linear model with a circular cross-section of radius $R_j$ in the jet, so that the axial periodicity in the linear model corresponds to the azimuthal periodicity around the jet in the simulation, Fig. \ref{fig:CFI_c}, upper panel. Under this identification, the axial wavenumber $k$ maps to the azimuthal mode number $m$ through $k = m/R_j$, so that . 
Figure~\ref{fig:A_linear} shows the linear growth rate for Case~A\_0.1 ($\Gamma=10$, $\sigma=0.1$, upper profiles of Fig.~\ref{fig:2DD}) as a function of wavenumber. 
Normalizing $k$ with the radius of curvature $R_c$ (as done in the linear analysis) yields 
$k R_c = m (R_c/R_j)$.

Using the measured values $R_c = 15.28$ and $R_j = 0.08$, we obtain $R_j/R_c \approx 0.0053$. 
For the $m=1$ mode (which corresponds to the maximum wavelength $2\pi R_j$) this gives $k R_c \approx 191$. Modes with a lower values of  $k R_c$ have a wavelength that cannot fit around the jet circumpherence and correspond to the green stable region  in Fig.~\ref{fig:A_linear}.
The linear analysis further predicts a most unstable mode at $k_{\rm max} R_c \approx 1100$.
Thus, Case~A lies well within the unstable range, confirming that the CFI develops physically in the simulation.

The linear stability analysis yields dimensionless growth rates expressed as $\mathrm{Im}\,\omega \, R_c/c$, where $R_c$ is the curvature radius of the cylindrical shell. In our simulations, $R_c \approx 15 z_0$, and the natural time unit is $z_0/c$. For the azimuthal mode number $m=7-8$, which corresponds to the dominant unstable mode observed in the simulations, the maximum growth rate reaches $\mathrm{Im}\,\omega\, R_c/c \approx 13$. Translating this into simulation time units, the growth time is $t_{\mathrm{grow}} = 1/\mathrm{Im}\,\omega = R_c/(\mathrm{Im}\,\omega \, c) \approx 15 z_0 / (13 c) \approx 1.15 \, z_0/c$.  This timescale indicates that the centrifugal instability develops in a time slightly longer than a single code unit $z_0/c$. However, the geometry is more complex than a simple cylinder, as the jet is deformed and the flow is advected along the axis. Taking advection into account, the growth time can be translated into a growth length by multiplying with the flow speed (approximated by $c$), which is consistent with the width of the recollimation region  where the instability develops in the simulations. 

Further insights are provided by Fig.~\ref{fig:3Dmapsprofs}, which displays, at $t=40~t_{\mathrm{cr}}$, longitudinal maps of the Lorentz factor $\Gamma$, and total pressure $P_\text{tot}$ (left panel), along with a map and transverse cuts of  $\rho$ (right panel) at the position indicated by the dashed line in the left panel.   In the density plot we can observe the beginning of the development of characteristic finger-like structures, signatures of growing CFI with m=8, that in fact corresponds to the value in which the linear analysis predicts the maximum growth rate (see Fig. \ref{fig:A_linear}).

\begin{figure*}
\begin{center}
\includegraphics[width=0.8\linewidth]{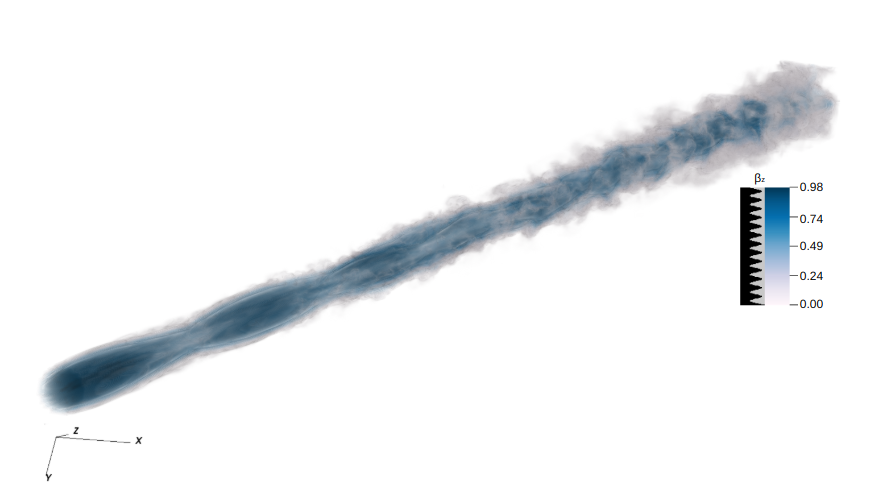}
\caption{3D volume rendering of the $z$-component of the jet 4-velocity ($\Gamma v_z$), case A\_0.1, $\sigma=0.1$ at t=80. The black-grey bar next to the colorbar indicates the opacity: grey regions correspond to opaque values of $\Gamma v_z$, while black regions indicate transparent regions.}
\label{fig:3D}
\end{center}
\end{figure*}

\begin{figure*}
\resizebox{\hsize}{!}{\includegraphics{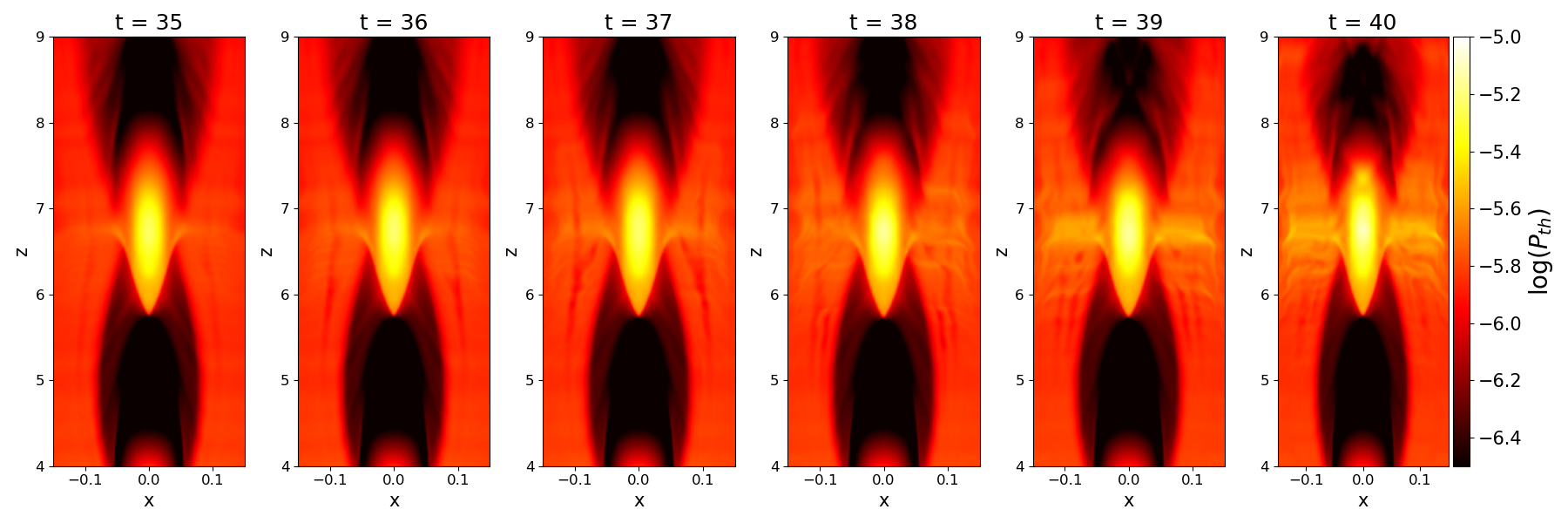}}
\caption{Time evolution of the thermal pressure distribution for case A\_0.1 ($\sigma = 0.1$) in the $x$--$z$ plane. The sequence of snapshots ($t=35$--$40$) shows the growth of the centrifugal instability: pressure perturbations at the jet edges gradually amplify, forming the characteristic lateral fingers and oscillatory structures along the jet axis. These features indicate the progressive destabilization of the jet due to centrifugal effects.}
\label{fig:cfi}
\end{figure*}

Considering that our simulations extend up to $t \sim 100\, z_0/c$, the observed rapid growth confirms that the instability fully develops and saturates early, in agreement with the nonlinear features and perturbations observed in the flow. Thus, the linear analysis provides insight into the onset and location of the centrifugal instability in our numerical models.
The large computational domain ($z \leq 30 z_0$) and long integration time ($t = 100\,t_\text{cr}$) allow us to track the whole evolution of the instability. Figure~\ref{fig:3D} shows a volume rendering of the $z$-component of the jet's 4-velocity ($\Gamma v_z$) of the 3D evolution of case A\_0.1, highlighting the emergence of turbulent structures. High-velocity regions and their interactions with surrounding, slower-moving material are clearly visible, indicating the onset of dynamic disruptions.  As shown in the pressure evolution of Fig.~\ref{fig:cfi}, the jet remains coherent and stable up to approximately $t \approx 40,t_\text{cr}$ (see also \citep{costa25} for a detailed discussion of the hydrodynamic case). Beyond this point, the growth of instabilities progressively disrupts the flow. In particular, the CFI continues to grow and is further amplified by the Richtmyer–Meshkov instability (RMI), leading to a coupling between the two modes \citep{Brouillette02,Zhou2021}. In plasmas, RMI develops when a shock wave crosses a perturbed interface between fluids of different densities, and in this context it reinforces the action of the CFI.

\begin{figure}[htbp]
\begin {center}
{\includegraphics[width=\linewidth]{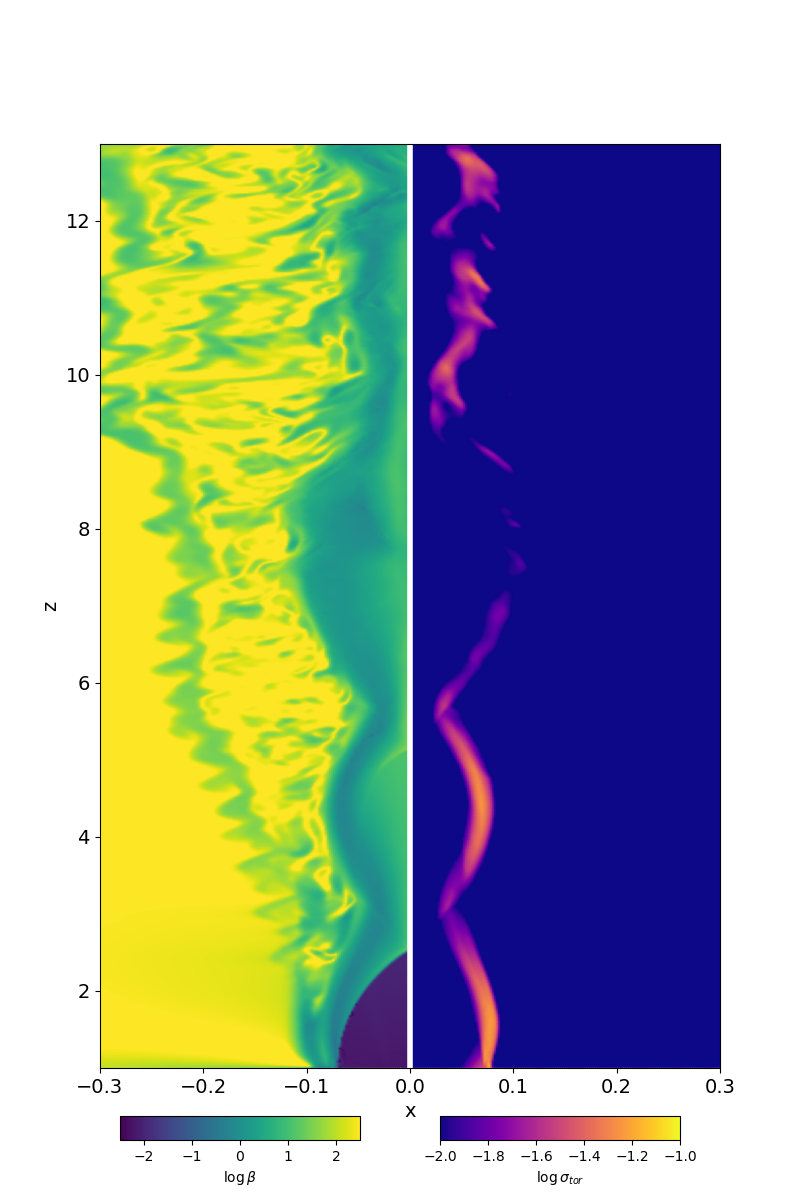}}
\caption{\footnotesize{Same as Fig.~\ref{fig:3Dmapsprofs}, but including MHD diagnostics (plasma $\beta$ on the left and the toroidal component of magnetization on the right) for the 3D case A\_0.1 with $\sigma = 0.1$.}}
\label{fig:3DMHD}
\end{center}
\end{figure}

Figure~\ref{fig:3DMHD} shows the plasma $\beta$ and the toroidal component of the magnetization $\sigma$ from the 3D simulation. These maps indicate that the jet remains coherent up to the second recollimation point ($z \approx 7 z_0$), beyond which it gradually decelerates and develops turbulence. The plasma $\beta$ provides a direct diagnostic of dissipation, distinguishing magnetically dominated regions (low $\beta$) from kinetically dominated ones (high $\beta$). In the early phase, the jet’s $\beta$ distribution is relatively uniform, while further downstream it exhibits strong fluctuations driven by nonlinear interactions. A central spine surrounded by a sheath becomes apparent, reflecting the impact of recollimation on the flow structure. The toroidal magnetization shown in the right panel emphasizes the role of the magnetic field in shaping this evolution, revealing how magnetic stresses contribute to both the stability and the eventual disruption of the jet.

We conclude with one final technical note. In starting a 3D calculation from the configuration produced by a 2D calculation, as we do here, one must be careful. As described in Appendix~\ref{app:2D3D}, the interpolation of the jet profile from a cylindrical grid to a Cartesian one at the injection boundary may introduce minor numerical artifacts. These arise from small mismatches at the boundary, which act as transient perturbations at the jet base. In marginally unstable cases, the combination of this initial perturbation and the underlying Cartesian symmetry can produce a weak cross-shaped pattern in the jet cross-section that derives from a preferential excitation of grid-aligned perturbations. In more unstable regimes, where physical modes grow faster, such numerical imprints are rapidly suppressed.  This perturbation, which we refer to as a “wave,” rapidly propagates up the jet axis, altering the jet structure. In magnetized cases, particularly those with higher $\sigma$, this transient wave can, in principle, seed asymmetric distortions of the magnetic field, potentially acting as a trigger for current-driven instabilities such as the kink mode further downstream. However, in the parameter regime explored here —where the focus is on the early development of the CFI— the influence of this perturbation remains limited. After the passage of the wave, the system relaxes back toward its initial equilibrium, and no sustained kink-like growth is observed within the simulated timescales. A more detailed analysis of the long-term evolution and possible onset of kink instabilities will be presented in a forthcoming work.

\subsubsection{A stable case}
As shown in Sect.~\ref{sec:2D}, Case~A\_0.1 provides insights into unstable configurations. Motivated by linear stability analysis for CFI, which suggests that jets with lower Lorentz factors are more stable, and by the results of \citet{Komissarov2019}, which indicate that smaller opening angles suppress instabilities, we select parameters that should favor a stable configuration. In this simulation the adopted parameters are: density ratio $\nu = 10^{-5}$, Lorentz factor $\Gamma = 5$, jet opening angle $\theta_\text{jet} = 0.05$, and pressure ratio $P_\text{ratio} = 10^{-3}$. 
\begin{figure}[htbp]
\centering
\includegraphics[width=\linewidth]{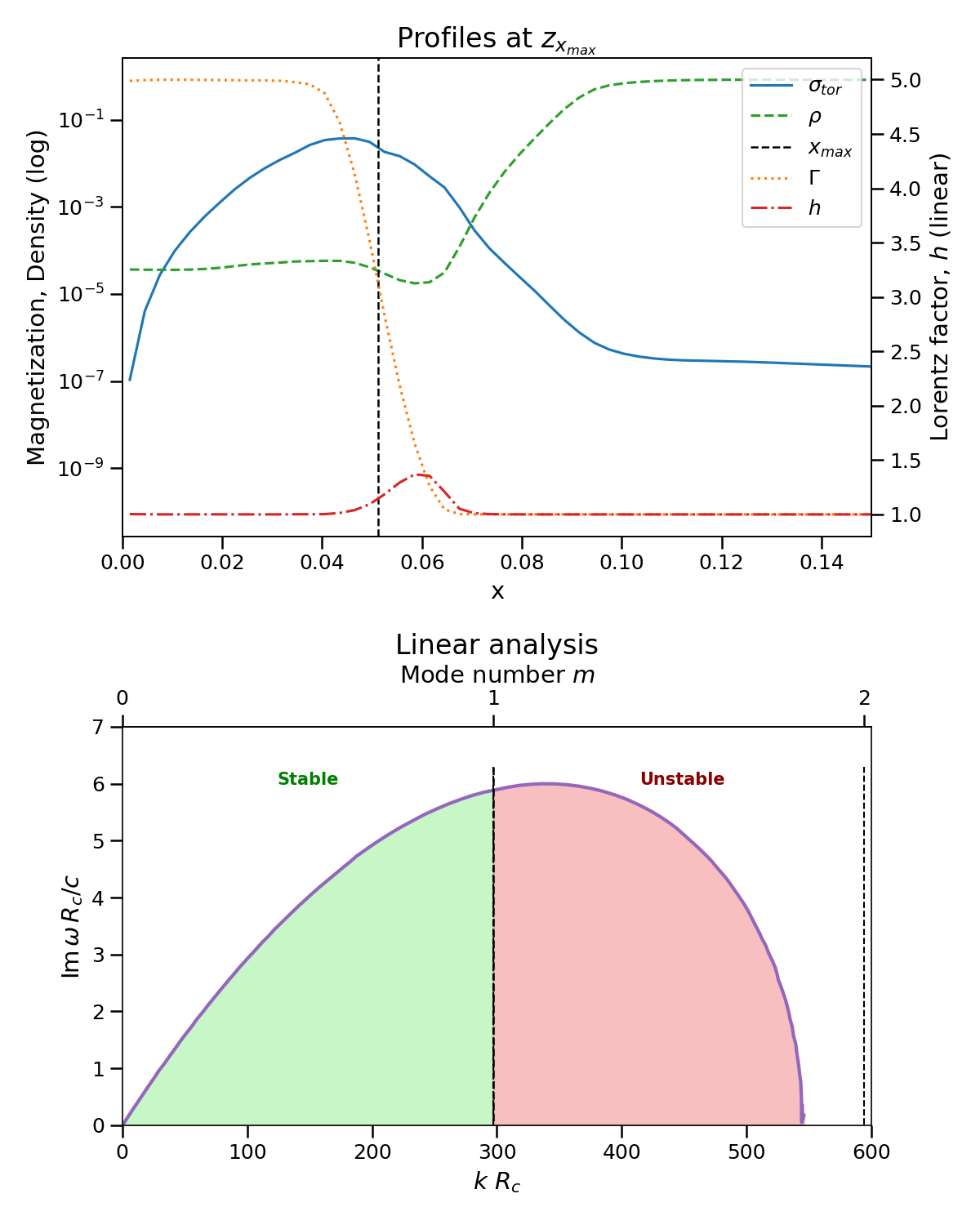}
   \caption{Top panel: Comparison of radial profiles of toroidal magnetization 
$\sigma_{\mathrm{tor}}$ (blue), Lorentz factor $\Gamma$ (orange), 
rest-mass density $\rho$ (green), and enthalpy $h/\rho$ (red) 
at $z = z_{\max}$Comparison of $\sigma_{tor}/\Gamma^2$ for jets with $\sigma=0.1$ with an opening angle $\theta=0.05$, Lorentz factor $\Gamma=5$. Bottom panel: Growth rate of the centrifugal instability (CFI) as a function of axial wavenumber $k ~R_c$. Shaded vertical bands indicate the values of $k ~R_c$ corresponding to azimuthal mode numbers $m = 1$ and $m = 2$ is, derived from the simulation jet radius $R_j = 0.0567$ and local curvature radius $R_c = 16.86$. The most unstable mode occurs at $k_{\text{max}} ~R_c \approx 300$.}
\label{fig:lin005}
\end{figure}

\begin{figure}[htbp]
\centering
\includegraphics[width=\linewidth]{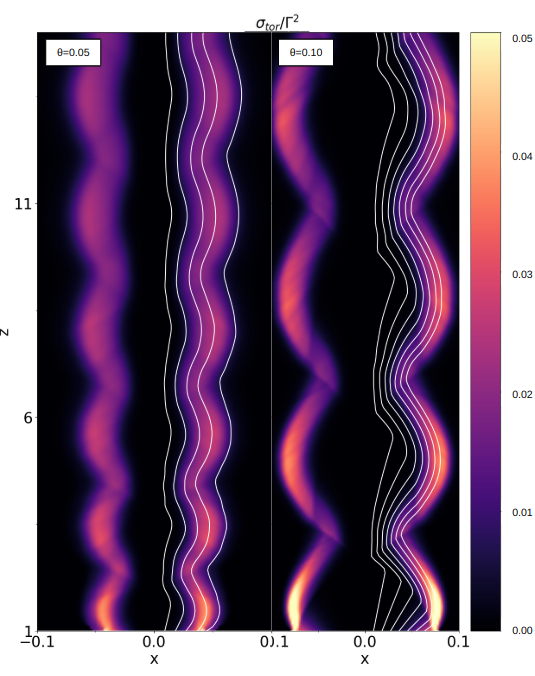}
   \caption{Comparison of $\sigma_{tor}/\Gamma^2$ for jets with $\sigma=0.1$ with a. opening angle $\theta=0.05$, Lorentz factor $\Gamma=5$ (left panel) and b. $\theta=0.1$, Lorentz factor $\Gamma=10$ (left panel) at $t=0$.}
\label{fig:3DMaa}
\end{figure}

\begin{figure}[htbp]
\centering
\includegraphics[width=0.95\linewidth]{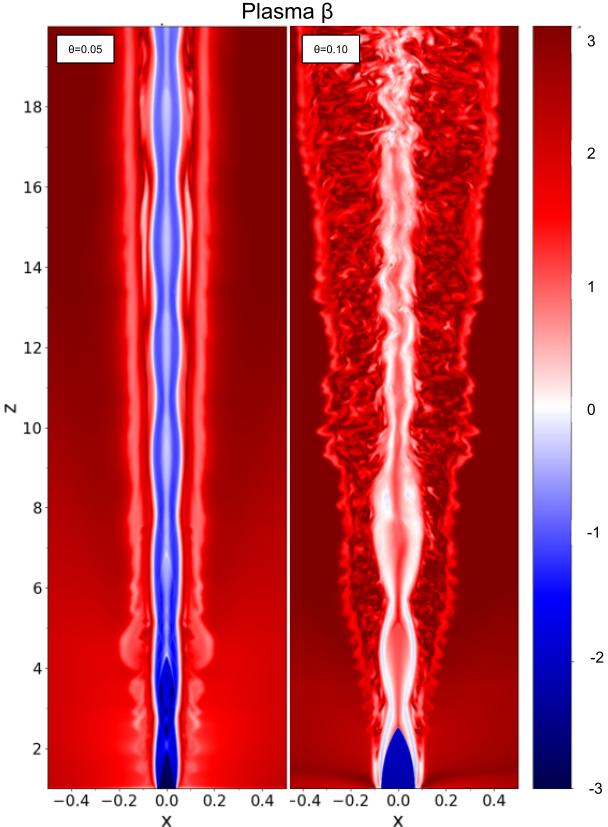}
   \caption{Comparison $\beta$ for jets with $\sigma=0.1$ with a. opening angle $\theta=0.05$, Lorentz factor $\Gamma=5$ (left panel) and b. $\theta=0.1$, Lorentz factor $\Gamma=10$ (left panel) at $t=80$.}
\label{fig:3Dcomp}
\end{figure}

The results confirm that the jet remains stable. To test the robustness of this conclusion, we extended the simulation up to $t = 100\,t_\text{cr}$ and found no transition to turbulence or structural deformation of the jet. 
The Lorentz factor remains steady along the axis, with no evidence of deceleration or anomalous pressure structures throughout the domain. An independent check based on the field geometry gives $\sigma_{\mathrm{tor}}/\Gamma^2 \approx 0.0523$ and $(\sigma_{\mathrm{tor}}/\Gamma)/(R_j/R) \approx 15.5 \gg 1$. 
This places the system well below the critical threshold for centrifugal instabilities, in agreement with the absence of unstable modes in the simulation. A posteriori, this evidence is confirmed by the linear analysis outlined in Sect.~\ref{sec:3DA}. 
The maximum growth rate occurs at $k_{\max} R_c \approx 350$.
With $R_c \approx 16.86$ and $R_j \approx 0.0567$, the resulting ratio $R_j/R_c \approx 3.4 \times 10^{-3}$ falls below the numerical threshold ($\sim 0.004$) required for the $m=4$ mode to develop, confirming that the configuration is linearly stable, as the linear analysis shows in Fig.~\ref{fig:lin005}. As shown in Fig.~\ref{fig:3DMaa} and~\ref{fig:3Dcomp}, the jet evolves smoothly without signs of disruption or turbulence. 

These parameters are in the stable regime and we further verified stability by increasing the jet density (i.e., varying $\nu$ to higher values; see Appendix~\ref{App:3}). The jet remains collimated and stable across these different setups.
This indicates that suppression of CFI in this parameter regime is robust and not sensitive to moderate changes in $\nu$.

The stabilizing effect is evident in Figs.~\ref{fig:3DMaa} and~\ref{fig:3Dcomp} (left panels), which show the distribution of $\sigma_{\text{tor}}/\Gamma^2$, streamlines, and plasma $\beta$. These diagnostics demonstrate that magnetic tension dominates over destabilizing forces, preventing the growth of unstable modes. A direct comparison with the unstable case underscores the crucial role of geometry and magnetization in controlling jet stability.   Together, these results show that narrower opening angles and stronger magnetic support shift the system away from the CFI threshold, enabling the jet to maintain coherence over large distances.

\section{Discussion and conclusions}\label{sec:4}

In this work, we examine how the addition of a magnetic field influences the stability of relativistic jets launched into an external medium with a vertical pressure profile that causes the jet to recollimate. In two-dimensional simulations, where axial symmetry is enforced, a stable sequence of recollimation shocks develops and the jet flow remains ordered. In a more realistic three-dimensional calculation,  non-axisymmetric modes and instabilities are now allowed to develop and may disrupt the jet. Indeed, earlier hydrodynamic calculations, e.g., \cite{costa25}, show that the curvature of the jet fluid flow lines that near a recollimation shock often enables the development and significant growth of the centrifugal instability (CFI). This in turn can lead to very efficient disruption of the jet flow after just one or two recollimation shocks, which may be relevant to the structure of jets observed in FRI and FR0 
radio sources. 

The addition of an internal jet magnetic field to the problem of a recollimating jet can have three important effects on the subsequent evolution of the jet. First, as we demonstrate in Sect. \ref{sec:2}, the addition of a magnetic field changes the vertical pressure profile of the jet, which in turn changes the position of the recollimation shocks and the curvature of the flow lines near the shocks. This quantitatively changes the values of the jet injection parameters that lead to a CFI-unstable flow.  Second, to the extent that fluid flow lines and magnetic field lines are tied together in the MHD approximation, the development of the CFI "corrugates" and stretches the magnetic field lines, increasing the effective tension of the field lines and thus resisting the further development of the CFI. A sufficiently strong field may thus quench the CFI and allow the jet propagate in a stable manner. Third, for sufficiently strong fields, other instabilities such as the current-driven instability (CDI) eventually become important and may again disrupt the jet. In this study, we restrict ourselves to the case of low to moderate jet magnetizations ($\sigma \sim 0-1$), where the CFI can still operate. For such magnetizations, our 3D results indicate that the CFI still seems to be the fastest growing initial instability, and thus the one relevant for determining where the jet flow eventually becomes unstable. Furthermore, we present a case that remains stable without triggering the CFI, indicating that the underlying stability problem is more complex than it may initially appear.

This conclusion is consistent with \citet{Matsumoto2021}, who also emphasized the relevance of the CFI in magnetized flows.  However, we find that the problem of determining exactly which initial jet parameters will lead to unstable jet flow is not as straightforward as implied in \citet{Komissarov2019}. For instance, a 3D case in \citet{Matsumoto2021} considered stable by the \citet{Komissarov2019} criterion becomes unstable when extended to a larger domain and longer runtime.  A further complication is that the nonlinear evolution of the jet after the CFI gets started appears to be non-trivial. In our 3D calculations of the jet flow evolution, the CFI appears to trigger other instabilities, and as in the hydrodynamic calculation of \cite{costa25}, different regions of the jet can eventually influence each other, possibly leading to mixing of jet material with the external medium and significantly changing the pressure profile of the medium near jet (as in the right panel of Fig. \ref{fig:3Dcomp}).  Again, one will not see this long-term behavior if the simulation is truncated in space or time. Conversely, if one sees an intrinsically young jet system in nature that does not show evidence (yet) of disruption, one should not conclude that the jet parameters are such that the jet is stable in the long-term.  

Our 2D survey shows that the location of the first recollimation shock is sensitive to the choice of initial jet parameters, namely the density contrast $\nu$, the pressure ratio $P_\text{ratio}$, the jet magnetization and the relative strength of the toroidal and poloidal field components, the Lorentz factor $\Gamma$, and the jet opening angle $\theta_\text{jet}$. Increasing the value of the non-magnetic field parameters tends to push the recollimation point to larger distances, modifying the jet's geometry. Physically, this reflects the enhanced inertial and pressure content of the flow, which requires more expansion and interaction with the external medium before reconfinement occurs. By contrast, higher magnetization moves the recollimation shock closer to the source. These trends may offer a valuable guide for interpreting observed structures in AGN jets, where knot positions are often correlated with recollimation features.

In summary, magnetization is a key factor in controlling jet collimation and the development of instabilities in the jet. While 2D studies and linear theory offer important guidance, only 3D simulations capture the full nonlinear evolution. These are currently computationally expensive to run for long enough times and with sufficient spatial resolution in a sufficiently large domain. We have therefore presented only a few illustrative 3D calculations, and further exploration of the full 3D problem is warranted.

Our results have direct implications for understanding the structure of AGN jets, the origin of knots and variability, and their polarized emission signatures.  For example, the development of the CFI-driven turbulence and magnetic field disordering could strongly influence the observed polarized emission \citep{sciaccaluga25}. In particular, the transition from a magnetically dominated spine to a turbulent sheath, or the radial mixing induced by recollimation and instability growth, can produce variable polarization degrees and swings. This is especially relevant in light of recent polarimetric observations from instruments like IXPE. Future work should focus on connecting the internal instability-driven structure of the jet with synthetic polarization maps to explore these effects in more detail.

\begin{acknowledgements}
We thank the anonymous referee for their useful comments. We thank Brian Reville, Konstantinos Nektarios Gourgouliatos, Dimitrios Millas, Paola Rossi and Om Sharan Salafia for fruitful discussions—special thanks to Om for Fig. \ref{fig:CFI_c}—and SB acknowledges helpful manuscript feedback from Brian and Dimitrios. We gratefully acknowledge financial support by INAF Theory Grant 2022 (PI F. Tavecchio) and NASA IXPE Theory Grant80NSSC24K1177 (P.I. P. Coppi).  This work was also funded by the European Union-Next Generation EU, PRIN 2022 RFF M4C21.1 (2022C9TNNX). We acknowledge support by CINECA, through ISCRA and Accordo Quadro INAF-CINECA, and by PLEIADI, INAF – USC VIII, for the availability of HPC resources (PI S. Boula). G.B. and A.C. acknowledge the support by the Spoke-1 "FutureHPC \& BigData” of the ICSC – Centro Nazionale di Ricerca in High Performance Computing, Big Data and Quantum Computing – funded by European Union – NextGenerationEU.
We have used the following Python libraries: Numpy \citep{numpy}, Matplotlib \citep{matplotlib}, Scipy \citep{scipy}, and PyPluto \cite{Pypluto25}.
\end{acknowledgements}

\bibliographystyle{aa}
\bibliography{refs} 

\begin{appendix}
\section{Smoothed profiles and initial conditions}\label{app:smooth}

The simulations adopt smoothed radial profiles to ensure a continuous transition between the jet and ambient medium, thereby minimizing numerical artifacts at the jet boundary. These profiles are implemented using the hyperbolic secant ($\operatorname{sech}$) function, which allows for smooth yet steep gradients.

The general expression for a smoothed quantity $q$ is given by:
\begin{equation}
q(r, z) = q_{\text{ext}} + (q_{\text{j}} - q_{\text{ext}})\, \operatorname{sech}\left[\left(\frac{r}{z \theta_q}\right)^{\alpha_q}\right],
\end{equation}
where $q_{\text{ext}}$ and $q_{\text{j}}$ are the values of the quantity in the ambient medium and jet, respectively; $\theta_q$ sets the angular width of the transition layer, and $\alpha_q$ controls the steepness of the profile.

The cylindrical radial distance $r$ and spherical radius $r'$ are defined as:
\begin{align}
r &= \sqrt{x^2 + y^2}, \\
r' &= \sqrt{x^2 + y^2 + z^2}.
\end{align}

The Lorentz factor profile is initialized as:
\begin{equation}
\Gamma(r, z) = 1 + (\Gamma_j - 1)\, \operatorname{sech}\left[\left(\frac{r}{z \theta_\Gamma}\right)^{\alpha_\Gamma}\right],
\end{equation}
from which the 3-velocity magnitude is calculated using:
\begin{equation}
v_0 = \sqrt{1 - \frac{1}{\Gamma^2}}.
\end{equation}

The density profiles are given by:
\begin{align}
\rho_{\text{ext}}(z) &= z^{-\eta}, \\
\rho_j(r') &= \nu\, r'^{-2}, \\
\rho_f(r, z) &= \rho_{\text{ext}} + (\rho_j - \rho_{\text{ext}})\, \operatorname{sech}\left[\left(\frac{r}{z \theta_\rho}\right)^{\alpha_\rho}\right],
\end{align}
where $\nu$ is the density contrast between the jet and the ambient medium, and $\eta$ controls the stratification of the external atmosphere.

Similarly, the pressure profiles follow:
\begin{align}
p_{\text{ext}}(z) &= P_0\, z^{-\eta}, \\
p_j(r') &= P_0\, r'^{-2\gamma}\, P_{\text{ratio}}, \\
p_f(r, z) &= p_{\text{ext}} + (p_j - p_{\text{ext}})\, \operatorname{sech}\left[\left(\frac{r}{z \theta_p}\right)^{\alpha_p}\right],
\end{align}
with $P_{\text{ratio}}$ denoting the jet-to-ambient pressure ratio.

For points inside the jet cone, defined by $r/z < \theta_j$, the velocity components are initialized as:
\begin{align}
u_x = v_0\, \frac{x}{r'}, ~
u_y = v_0\, \frac{y}{r'}, ~
u_z = v_0\, \frac{z}{r'}.
\end{align}

The parameters used to define the smoothed profiles and physical conditions in the simulations are: $\theta_j = 0.1$, $\Gamma_j = 10$, $P_0 = 3 \times 10^{-6}$, $\theta_\Gamma = 0.07$, $\alpha_\Gamma = 13$, $\theta_p = 0.13$, $\alpha_p = 15$, $\theta_\rho = 0.1$, $\alpha_\rho = 15$, $\theta_\beta = 0.004$, $b_z = 10^{-5}$, and $\alpha_z = 10^{-10}$.

\section{Resolution of the simulations}\label{app:3dres}
The 2D simulations use a grid of $1400 \times 2200$ zones. In the central region  $[0, 2] \times [1, 20]$, the grid is uniform, providing a resolution of 50 points per  initial jet radius, while the outer regions are geometrically stretched. The X1-grid  comprises two regions: an inner uniform region $r \in [0, 1000]$ with 400 zones, and  an outer stretched region with  up to 6. Similarly, the X2-grid has two regions: 
an inner uniform region $z \in [1, 1500]$ with 700 zones, and an outer stretched region up to 30. At the final time, the 2D simulations reach a nearly steady state, which is used as the initial condition  for the 3D runs.

The 3D simulations are performed on a Cartesian domain with coordinates 
\[
x \in \left[-\frac{L_{x,3D}}{2}, \frac{L_{x,3D}}{2}\right], \quad
y \in \left[-\frac{L_{y,3D}}{2}, \frac{L_{y,3D}}{2}\right], \quad
z \in [1, L_{z,3D}],
\]
where lengths are expressed in units of $z_0$, $L_x=L_y$, and $z$ is the jet propagation direction. 
A uniform grid spacing is adopted along $z$ in the central jet region, while stretched grids are used 
in the transverse directions and in the far axial domain. Table~\ref{tab:3d_grids} summarizes the 
resolution, grid structure, and number of cells per initial jet radius for the cases discussed in 
Sec.~\ref{sec:3D} (A1 and B1) and in Appendix~\ref{App:3} (A2 and B2).

\begin{table*}[h]
\centering
\label{tab:3d_grids}
\scriptsize
\begin{tabularx}{\textwidth}{l c c c X X c}
\hline\hline
Simulation & Parameters $(\theta_j, \sigma, \nu)$ & $N_x \times N_y \times N_z$ & Domain $(L_x,L_y,L_z)$ & $x,y$ grid & $z$ grid & Cells per $R_{j,0}$ \\
\hline
A1 & $(0.1, 0.1, 10^{-5})$ & $701 \times 701 \times 1000$ & $[-2,2]\times[-2,2]\times[1,20]$ & 
stretched in $[-2,-0.5]$ and $[0.5,2]$, uniform in $[-0.5,0.5]$ & uniform (1--20, 1000 u) & 50 \\

A2 & $(0.1, 0.001, 10^{-5})$ & $257 \times 257 \times 1024$ & $[-0.15,0.15]\times[-0.15,0.15]\times[1,6]$ & 
uniform ($-0.15$--0.15) & uniform (1--6, 1024 u) & 85 \\

B1 & $(0.05, 0.1, 10^{-5})$ & $501 \times 501 \times (800+350)$ & $[-2.5,2.5]\times[-2.5,2.5]\times[1,30]$ & 
stretched in $[-2.5,-0.3]$ and $[0.3,2.5]$, uniform in $[-0.3,0.3]$ & uniform (1--20, 800 u), stretched (20--30, 350 s) & 30 \\

B2 & $(0.05, 0.1, 10^{-4})$ & $501 \times 501 \times (2000+350)$ & $[-2.5,2.5]\times[-2.5,2.5]\times[1,40]$ & 
stretched in $[-2.5,-0.3]$ and $[0.3,2.5]$, uniform in $[-0.3,0.3]$ & uniform (1--30, 1200 u), stretched (30--40, 350 s) & 30 \\
\hline
\end{tabularx}
\caption{Grid setup of 3D simulations.}
\end{table*}

\section{Comparison 2D and 3D}\label{app:2D3D}
\begin{figure}
\centering
\includegraphics[width=0.5\linewidth]{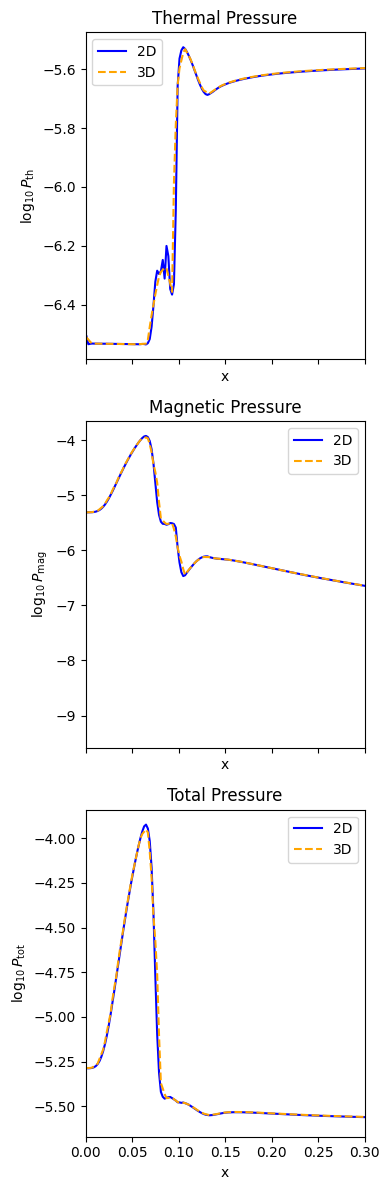}
\caption{Comparison of thermal, magnetic, and total pressure profiles along the x-axis at 
$z=z_0$, for the 2D simulation at t=3000 and the 3D simulation at t=0. The 3D data have been interpolated onto the 2D spatial grid for a consistent comparison. }
\label{fig:2D3D}
\end{figure}

To initialize the 3D simulations from 2D axisymmetric profiles, we employed the \texttt{InputDataInterpolate} and \texttt{StaggeredRemap} utility functions provided by PLUTO. These tools ensure consistent remapping of primitive and staggered quantities from 2D cylindrical coordinates to the 3D Cartesian grid. Figure~\ref{fig:2D3D} shows a comparison of the pressure profiles—thermal, magnetic, and total—evaluated at the same axial position $z=z_0$. The agreement between the 2D and 3D profiles confirms the accuracy of the interpolation procedure. Minor discrepancies appear in the thermal pressure, particularly for $x<0.1$, where the 3D profile is slightly smoother. This is due to the interpolation onto a lower-resolution 3D grid, which tends to average out sharp features. Nonetheless, the essential physical structure is preserved, validating the initialization approach.

\begin{figure}
\resizebox{\hsize}{!}
{\includegraphics{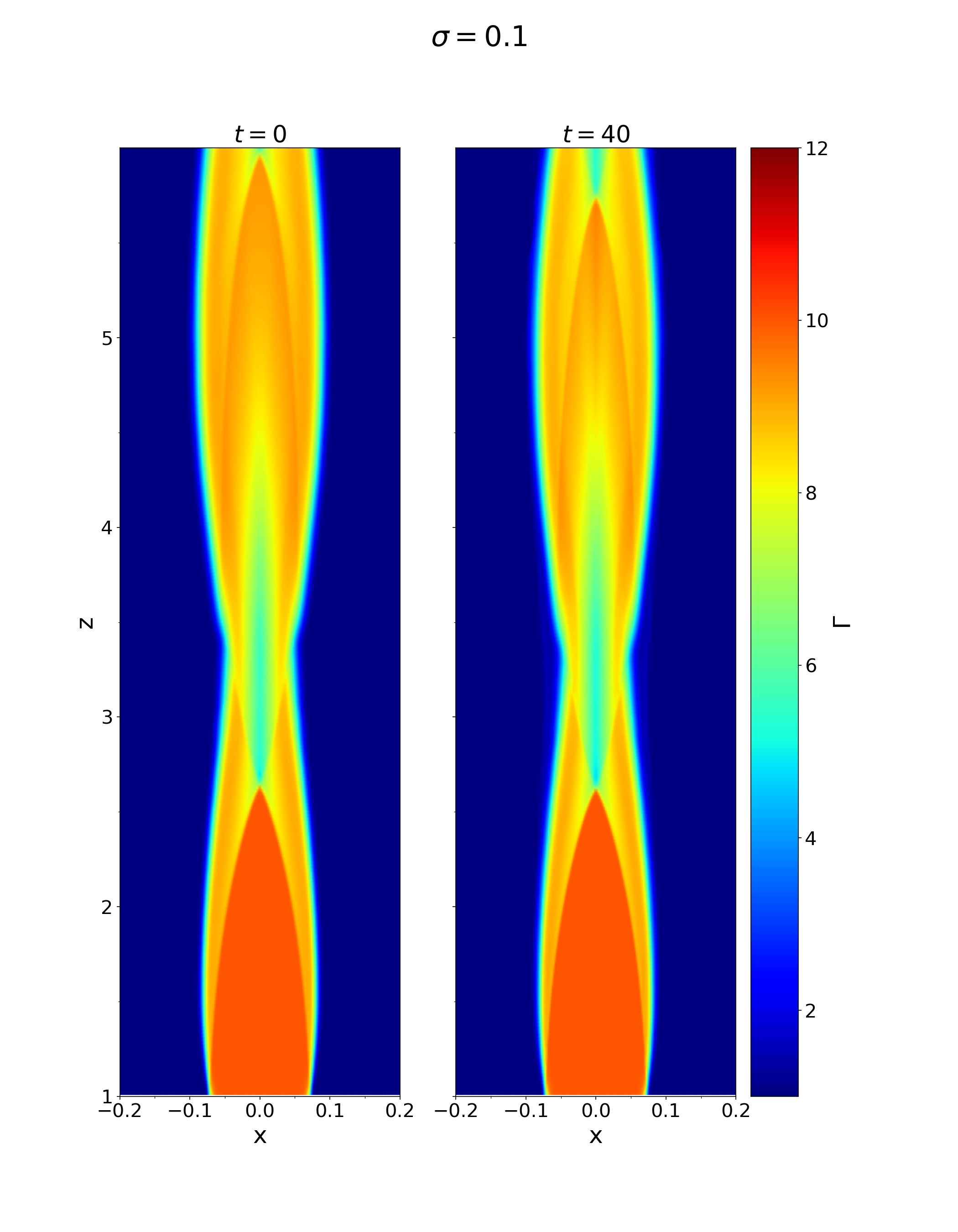}}
\caption{Comparison of the Lorentz factor at $t = 0$ and $t = 40$ for case A\_0.1 with $\sigma = 0.1$, illustrating the jet structure before and after the impact of the imposed wave. }
\label{fig:3Dwave}
\end{figure}

Figure~\ref{fig:3Dwave} shows the Lorentz factor at $t = 0$ and $t = 40$, illustrating that the first recollimation point remains unaffected by the initial numerical wave, which acts as a perturbation. The observed differences at the second recollimation point are instead due to the physical evolution of the system, as this is the time when the instability begins to develop.

\section{3D simulations-other parameters}\label{App:3}
\begin{figure*}
\centering
\resizebox{0.925\hsize}{!}{\includegraphics[clip=true]{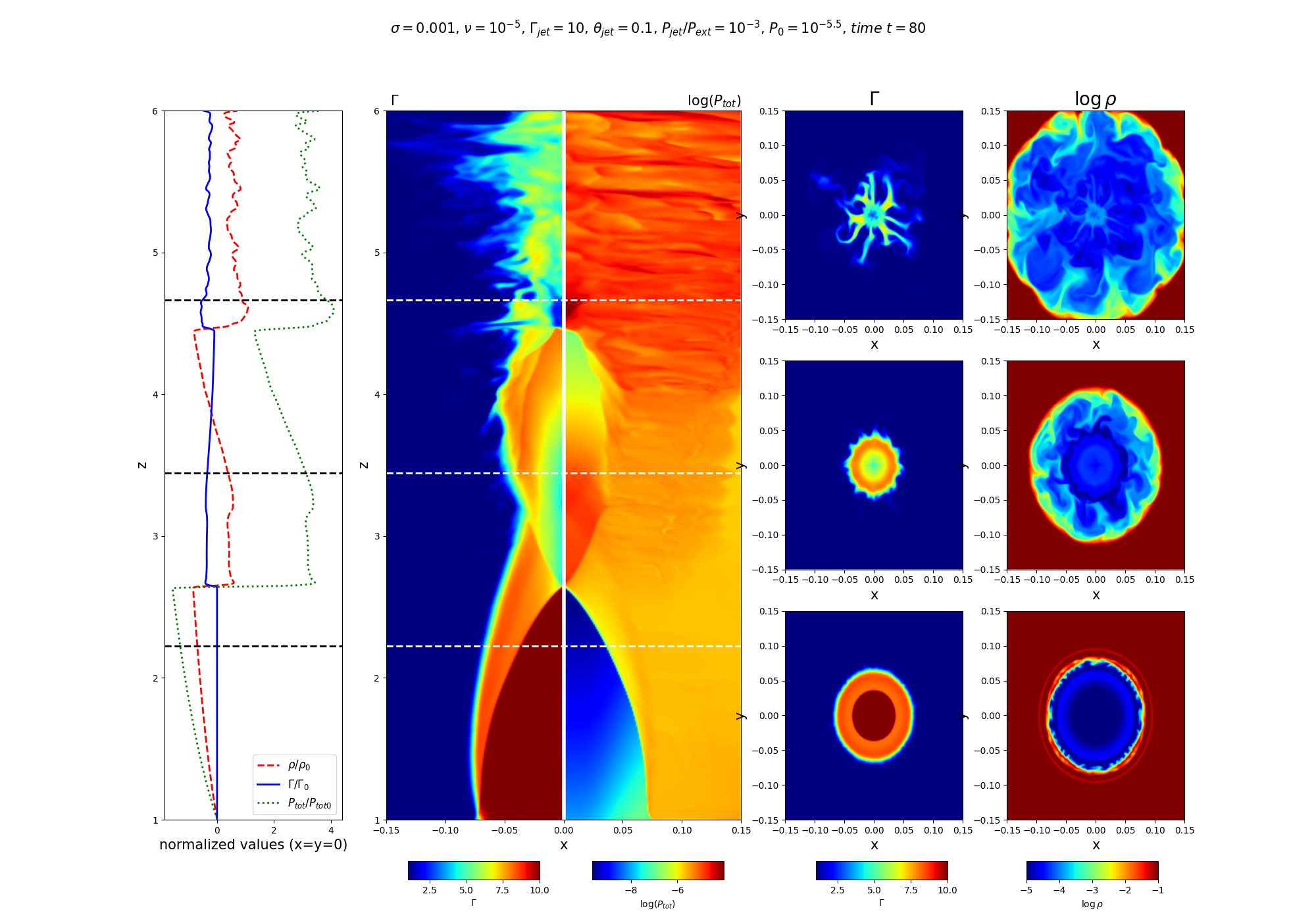}}
\caption{\footnotesize{For the 3D case A with magnetization $\sigma = 0.001$, the plot in the first column shows the logarithmic normalized profiles along the $z$-axis at position $x=0$ for density $\rho$, Lorentz factor $\Gamma$, and total pressure $P_\text{tot}$. The other columns include the map of $\Gamma$ and $\log P_\text{tot}$, and cuts at three different $z$-positions, where the lines in the profiles and maps correspond to these cuts, showing the Lorentz factor and logarithmic density.}}
\label{fig:sigma0001}
\end{figure*}

\begin{figure*}
\centering
\resizebox{0.925\hsize}{!}{\includegraphics[clip=true]{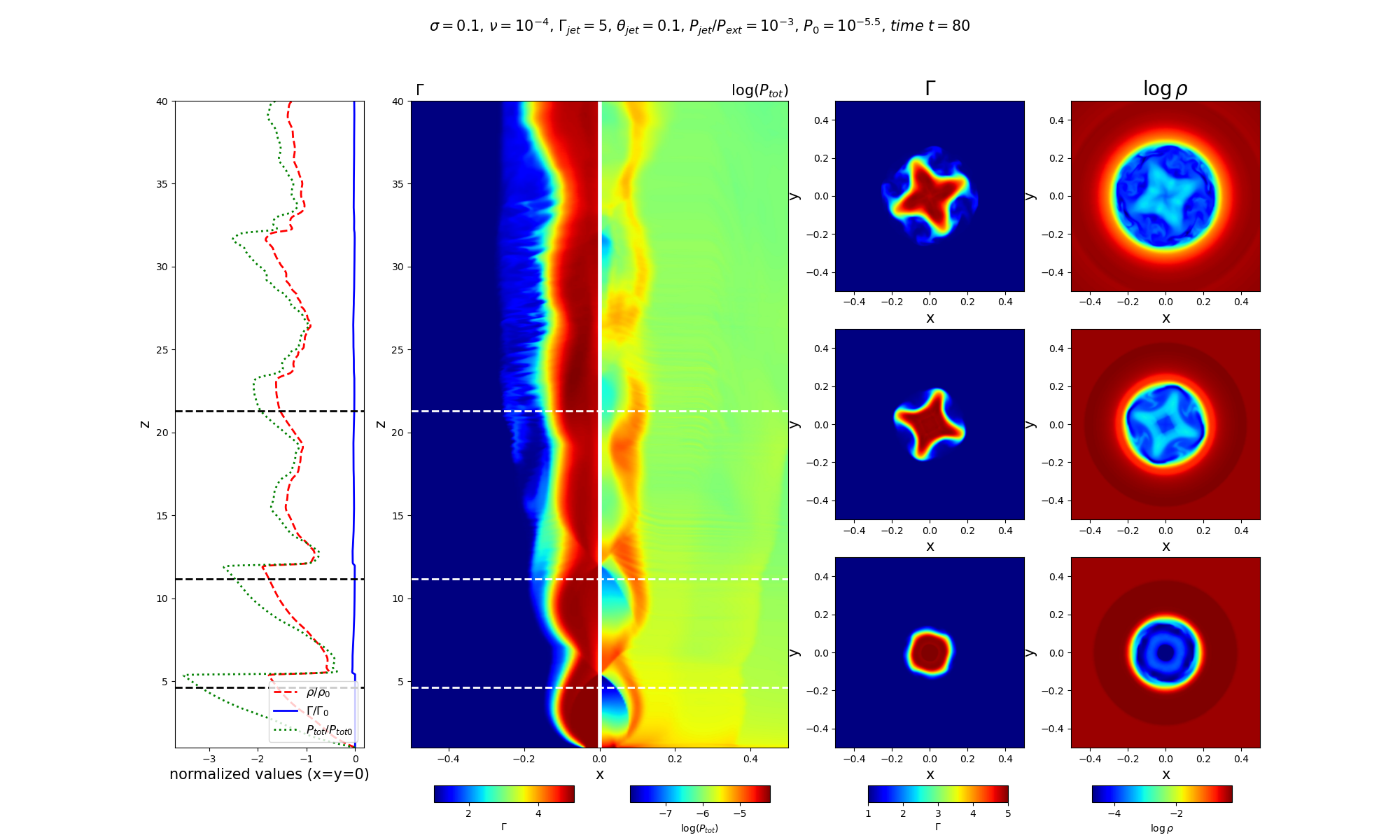}}
\caption{\footnotesize{Same as in Fig. \ref{fig:sigma0001}. For the 3D case with magnetization $\sigma = 0.1$, $\theta_{jet}=0.05$, $\nu=10^{-4}$ and $\Gamma=5$.}}
\label{fig:005heavy}
\end{figure*}

The appendix provides additional diagnostics of the jet evolution. We show logarithmic normalized profiles along the $z$-axis at $x=0$ for the density $\rho$, Lorentz factor $\Gamma$, and total pressure $P_\text{tot}$, together with maps and transverse cuts of $\Gamma$ and $\log P_\text{tot}$ at several positions along the jet. These allow us to compare the longitudinal evolution of the main dynamical quantities with the corresponding cross-sectional structure, highlighting differences between unstable and stable configurations.
Figures~\ref{fig:sigma0001} and \ref{fig:005heavy} show the evolution of two jets with different initial parameters. In Fig.~\ref{fig:sigma0001}, the jet is initialized with magnetization $\sigma = 10^{-3}$, density ratio $\nu = 10^{-5}$, Lorentz factor $\Gamma = 10$, opening angle $\theta_\text{jet} = 0.1$, and pressure ratio $P_\text{ratio} = 10^{-3}$. This configuration leads to a rapid destabilization, closely resembling the behavior seen in hydrodynamic jets. The structure breaks down early in the simulation, with no clear recollimation features, indicating that the magnetic support is insufficient to suppress growing instabilities.

By contrast, the jet in Fig.~\ref{fig:005heavy}, though having $\sigma = 10^{-1}$, is initialized with a higher density ratio $\nu = 10^{-4}$, a lower Lorentz factor $\Gamma = 5$, and a smaller opening angle $\theta_\text{jet} = 0.05$. This jet remains stable over a longer distance and preserves its structure. Longitudinal profiles show smoother trends in $\rho$, $\Gamma$, and $P_\text{tot}$, while transverse cuts of $\Gamma$ and $\log P_\text{tot}$ reveal only mild deformations. These distortions are not associated with CFI, but rather suggest the presence of weaker or subdominant modes. In the limit where the jet remains stable, the cuts reveal a weak but coherent $m=4$ signature, likely related to the underlying grid symmetry. However, this grid-induced mode is naturally enhanced due to symmetry and grows faster than physical instabilities, such as the CFI. Although present, it does not affect the overall stability or evolution of the jet. The increased stability is attributed not only to the slightly higher effective magnetization 
but also to the narrower opening angle, which reduces the jet's initial expansion and limits the triggering of instabilities. This highlights how geometric and dynamical parameters jointly influence the onset and growth of jet instabilities.

The effect of the imposed wave perturbation also differs significantly between these two cases. In the stable jet configuration (Fig.~\ref{fig:005heavy}), the wave acts as a mild perturbation rather than a destabilizing driver. Here, the jet's conditions effectively suppress the CFI, and the wave does not significantly amplify instability growth. However, grid-related numerical instabilities can still manifest, which explains the appearance of four discrete modes observed in the transverse cuts. These modes correspond to perturbations seeded by the numerical grid rather than physical instabilities, and crucially, they do not disrupt the jet's overall stability. Additionally, the presence of jet rotation induces a kink-like deformation, visible in the cuts; however, this kink instability remains benign and does not interrupt the jet flow.
Conversely, in the unstable jet case (Fig.~\ref{fig:sigma0001}), the jet is inherently prone to CFI due to its initial conditions. Here, the imposed wave perturbation has little to no influence on the jet's evolution; the CFI dominates and drives the rapid destabilization observed. The intrinsic CFI dominates the jet dynamics, rendering the wave perturbation negligible in driving the disruption of the jet structure.

These results highlight the intricate interplay between physical instabilities and perturbations. While waves can seed or amplify instabilities in marginally stable jets, they are insufficient to destabilize strongly unstable configurations dominated by current-driven modes.


\end{appendix}

\end{document}